\documentclass[a4paper]{article}

\usepackage{amssymb}
\usepackage{amsmath}
\usepackage{mathrsfs} 
\usepackage[all]{xy}
\usepackage[dvips]{color}
\usepackage[dvips]{graphicx}

\definecolor{dyellow}{rgb}{1.,0.8,.0}
\definecolor{myblue}{rgb}{.1,.1,.7}
\definecolor{dcyan}{rgb}{.0,.6,.6}
\definecolor{dmagenta}{rgb}{0.6,0.0,0.6}
\definecolor{brown}{rgb}{0.6,0.2,0.}
\definecolor{darkblue}{rgb}{.0,.0,0.5}
\definecolor{darkred}{rgb}{0.75,0.0,0.0}
\definecolor{orange}{rgb}{1.,.6,.0}
\definecolor{dorange}{rgb}{0.8,.4,.0}
\definecolor{darkgreen}{rgb}{0.0,0.6,0.0}
\definecolor{purple}{rgb}{.4,.0,.4}
\definecolor{grey}{rgb}{0.7,0.7,0.7}

\newcommand{\hide}[1]{}
\newcommand{\delete}[1]{}
\newcommand{\distribute}[3]{} 

\newcommand{\vect}[1]{\ensuremath{\boldsymbol{#1}}}
\newcommand{\tensor}[1]{\mathbf{#1}}

\newcommand{\diag}{\mathrm{diag}}

\newcommand{\mathspan}{\mathrm{span}}

\newcommand{\od}{\mathrm{d}}

\newcommand{\at}[1]{\big|_{#1}}

\newcommand{\mathmatrix}[2]{
  \left(
    \begin{array}{#1}
      #2
    \end{array}
  \right)
}

\title{Smooth Solutions and Discrete Imaginary Mass \\
  of the Klein-Gordon Equation in the de~Sitter Background}
\author{Bin Zhou\thanks{Email: \texttt{zhoub@bnu.edu.cn}} and
  Zhen-Hua Zhou\thanks{Email: \texttt{dtplanck@163.com}} \\
  Department of Physics, Beijing Normal University, Beijing 100875, P. R. China}
\date{December 26, 2011}

\begin{document}
\maketitle

\begin{abstract}
Using methods in the theory of semisimple Lie algebras, we can obtain all smooth
solutions of the Klein-Gordon equation on the 4-dimensional de~Sitter spacetime
($dS^4$).  The mass of a Klein-Gordon scalar on $dS^4$ is related to
an eigenvalue of the Casimir operator of $\mathfrak{so}(1, 4)$.
Thus it is discrete, or quantized.
Furthermore, the mass $m$ of a Klein-Gordon scalar on $dS^4$ is imaginary:
$m^2 \propto - N (N + 3)$, with $N \geqslant 0$ an integer.
\end{abstract}


\section{Introduction}
\label{sect:intro}

Ever since 1900, quantum theories (quantum mechanics and quantum field theory)
and the theories of relativity (special and general relativity) are the most
significant achievements in physics.  However, for more than 100 years,
the compatiblity of these two categories of theories is still a problem.
Needless to say the conflict in spirit, only the obstacle in technique passing
from general relativity to quantum theory has been enough problematic.

In the process of seeking for a theory of quantum gravity, there is the effort
to establish QFT in curved spacetime\cite{Kay, Wald06, Hollands-Wald-0803,
Hollands-Wald-0805, Ind-Lib, Wald2009}, assuming that the gravitational fields
are dominated by a classical theory (typically by general relativity).
The classical background may or may not be affected by the quantum fields.
Currently the QFT in curved spacetime is mainly a framework generalized
from the QFT in the Minkowski spacetime.
Difference between a generic curved spacetime and the Minkowski spacetime
has been considered, but, without much detailed knowledge of classical solutions
of field theories in curved spacetime, some key points in QFT in curved
spacetimes are inevitable questionable.

For a similar example, let us examine what happens when we consider
scalar fields in $(1 + 1)$-dimensional toy spacetimes.
If the spacetime is a Minkowski spacetime, scalar fields on it can be analyzed
using the Fourier transform with respect to the spatial direction.
If the space is closed as the circle $S^{1}$, however, scalar fields turn out
to be Fourier series with respect to the spatial direction.  In such an example,
we can see how the difference of topologies could make great influence on
the analysis tools.

Given a curved spacetime, its topological structure and geometric structure
may affect many aspects of the fields.  For a linear field equation, for
instance, whether its solution space is infinite dimensional, whether some
parameters (such as the mass) of the field equations are restricted to take
only some special values, and so on, all depend on these structures.
The QFT in curved spacetime must take these aspects into account.

In this paper, we take the Klein-Gordon equation on the 4-dimensional
de~Sitter spacetime (denoted by $dS^{4}$ in this paper) as an example.
Applying the representation theory of semisimple Lie algebras, we can obtain
\textit{all smooth classical solutions} of the Klein-Gordon equation
on $dS^{4}$.
It is clear that, unlike the case of the Minkowski spacetime, the solution space
of the K-G equation on $dS^{4}$ is finite dimensional.  What's more important
is that the mass of a Klein-Gordon equation on $dS^{4}$ cannot be continuously
real-valued.  In fact, there exists nonzero solutions when and only when
the mass $m$ satisfies
\begin{equation}
  m^{2} = - N (N + 3) \, \frac{\hbar^{2}}{c^{2} l^{2}}
  \label{eq:mass-spectrum}
\end{equation}
for certain a nonnegative integer $N$, with $l$ the ``radius'' of $dS^{4}$.
In the quantization process, the negative sign in $m^{2}$ should be a great
obstacle to interpret $m$ to be the mass of particles excited by quantum
scalar fields.

This paper is organized as follows.
In Section~\ref{sect:outline} we outline our idea of how to obtain all smooth
solutions of the Klein-Gordon equation on $dS^{4}$.  We start the outline from
the well know theory of angular momentum in quantum mechanics.
In Section~\ref{sect:prog} we describe the whole principle and program of
our method.
In Section~\ref{sect:Lie-alg} we apply the theory of semisimple Lie algebras
to $ \mathfrak{so}(1, 4) \otimes_{\mathbb{R}} \mathbb{C}$, the complexification
of $ \mathfrak{so}(1, 4) $, and describe its algebraic structures which have
nothing to do with its representations.
For convenience $ \mathfrak{so}(1, 4) \otimes_{\mathbb{R}} \mathbb{C}$
is denoted by $ \mathfrak{L} $.
In Section~\ref{sect:irred-modules} we apply the representation theory of
semisimple Lie algebras to the $ \mathfrak{L} $-module $ C^{\infty}(dS^{4}) $,
analogous to the coordinate representation of angular momentum operators
in quantum mechanics.
In this section we give some coordinate systems adapted to
the Cartan subalgebras of $ \mathfrak{L} $, and construct irreducible
$ \mathfrak{L} $-submodules of $ C^{\infty}(dS^{4}) $.
In Section~\ref{sect:solution-mass} we give the mass of the Klein-Gordon
equation and its smooth solutions.
Then, finally, in Section~\ref{sect:CD}, we give the summary and discuss some
related problems.
Detailed investigation and proofs are left in the appendices.

\section{Outline of the Ideas}
\label{sect:outline}

In general relativity, the Klein-Gordon equation reads
\begin{equation}
  g^{ab} \nabla_{a} \nabla_{b} \phi + \frac{m^2 c^2}{\hbar^2} \, \phi = 0
  \, ,
  \label{eq:KG}
\end{equation}
where $\nabla_{a}$ is the Levi-Civita connection. (Indices like $ a $ and $ b $
are abstract indices~\cite{Wald}.) In this paper the signature of
the metric tensor field $g_{ab}$ is like that of
\begin{equation}
  (\eta_{\mu\nu})_{4 \times 4} = \diag(1, -1, -1, -1)
  \, .
\end{equation}

The Klein-Gordon equation is a linear PDE.  To solve a PDE, one often applies
the method of separation of variables.  However, whether this method works 
depends heavily on the choice of the coordinates $x^\mu$.  If one is not so
lucky to choose the right coordinates, this method would fail even if the PDE
can be easily solved in certain a right coordinate system.

In order to solve a linear PDE using the method of separation of variables,
we must choose a coordinate system that is closely related to the symmetry of
the PDE.  When the symmetry group is a Lie group with a sufficient large rank
(the dimension of its Cartan subgroup), it is even possible to solve
the PDE by virtue of the theory of Lie groups and Lie algebras.
In this case the method of separation of variables is only needed to find out
a maximal vector\cite{Humphreys} corresponding to the highest weight.
After the maximal vectors have been found out, the solution space of the PDE
can be easily, but often tediously, constructed.

To show this, we take the equation
\begin{equation}
  h^{\alpha\beta} D_{\alpha} D_{\beta} Y = - \lambda Y
  \label{eigen-eq:Laplace:S2}
\end{equation}
on $S^{2}$ (the unit 2-sphere) as an example, where $ D_{\alpha}$ is
the Levi-Civita connection compatible with the standard metric tensor field
$ h_{\alpha\beta} $ on $ S^{2} $.  This is the eigenvalue equation of
the Laplacian operator for 0-forms on $S^{2}$.  Since the symmetry group
of $S^{2}$ is $O(3, \mathbb{R})$, and since eq.~(\ref{eigen-eq:Laplace:S2})
is determined by the metric tensor field $h_{\alpha\beta}$, the symmetry group
of eq.~(\ref{eigen-eq:Laplace:S2}) contains $O(3, \mathbb{R})$
as a Lie subgroup.  In the spherical coordinate system $(\theta, \varphi)$,
eq.~(\ref{eigen-eq:Laplace:S2}) takes the well known form
\begin{equation}
  \frac{1}{\sin\theta} \, \frac{\partial}{\partial \theta}
    \Big( \sin\theta \frac{\partial Y}{\partial \theta} \Big)
  + \frac{1}{\sin^2 \theta} \, \frac{\partial^2 Y}{\partial \varphi}
  = - \lambda \, Y
  \, .
  \label{eq:Laplace-S2}
\end{equation}

On the one hand, this PDE arises from the Helmholtz equation by
separation of variable, and a further separation of variable leads to
the general Legendre equation, together with the equation of
a simple harmonic oscillator with respect to the variable $\varphi$.
As a result, the solution of eq.~(\ref{eq:Laplace-S2}), namely,
eq.~(\ref{eigen-eq:Laplace:S2}),
is a linear combination
\begin{equation}
  Y = \sum_{m = -l}^l C_{m} \, Y_{lm}(\theta, \varphi)
  \label{eq:Y:o(3)}
\end{equation}
of spherical harmonics $Y_{lm}(\theta, \varphi)$ with a fixed nonnegative
integer $l$, where $C_{m}$ are arbitrary constants, and $\lambda = l (l + 1)$.

On the other hand, in quantum mechanics, there is a more elegant method
using the theory of angular momentum operators to solve
eq.~(\ref{eq:Laplace-S2}).  In geometry, the angular momentum operators
$\hat{\vect{L}} = \vect{r} \times \hat{\vect{p}}
  = -i \hbar \, \vect{r} \times \nabla$
are proportional to three Killing vector fields
\begin{alignat}{1}
  I_{x} &= \frac{1}{i \hbar} \, \hat{L}_{x}
  = \sin\varphi \, \frac{\partial}{\partial \theta}
  + \cot\theta \cos\varphi \, \frac{\partial}{\partial \varphi}
  \, ,
  \\
  I_{y} &= \frac{1}{i \hbar} \, \hat{L}_{y}
  = - \cos\varphi \, \frac{\partial}{\partial \theta}
  + \cot\theta \sin\varphi \, \frac{\partial}{\partial \varphi}
  \, ,
  \\
  I_{z} &= \frac{1}{i \hbar} \, \hat{L}_{z}
  = - \frac{\partial}{\partial \varphi}
  \, ,
  \label{eq:Killing-3:S2}
\end{alignat}
respectively.  To apply the theory of angular momentum operators
to the coordinate representation is, in fact, equivalent to apply
the representation theory of $\mathfrak{so}(3, \mathbb{R})$
to $C^{\infty}(S^{2})$, the $\mathfrak{so}(3, \mathbb{R})$-module of all smooth
functions on $S^{2}$.  In this manner $ Y_{ll}(\theta, \varphi) $, which acts
as the maximal vector in the representation theory, can be obtained by solving
two PDEs of order one:
\[
  \hat{L}_{z} Y_{ll} = l \hbar \, Y_{ll}
  \, , \qquad
  \hat{L}_{+} Y_{ll} = 0
  \, .
\]
Then all other $Y_{lm}(\theta, \varphi)$ can be obtained from
\[
  \hat{L}_{-} Y_{lm} = \hbar \sqrt{(l + m)(l - m + 1)} \, Y_{l, m - 1}
\]
for $ m = l $, $ l - 1 $, \ldots, $ -l $ recursively.
It is well known that eq.~(\ref{eq:Laplace-S2}), namely,
eq.~(\ref{eigen-eq:Laplace:S2}), is equivalent to
\[
  \hat{\vect{L}}^{2} Y = \lambda \hbar^{2} \, Y
  \, ,
\]
and $\hat{\vect{L}}^{2}$ is a Casimir operator of
$\mathfrak{so}(3, \mathbb{R})$.
Because all $Y_{lm}$ with $m = l$, $l - 1$, \ldots, $- l$ span an irreducible
$\mathfrak{so}(3, \mathbb{R})$-module, the linear combination (\ref{eq:Y:o(3)})
is automatically a solution of the above equation, according to
Schur's lemma~\cite{Humphreys}.  In order to determine the value of $\lambda$,
we need simply substitute $Y = Y_{ll}$ into the above equation, obtaining
$\lambda = l(l + 1)$.

The reason that separation of variables works is deeply related to the fact that
the $ \varphi $-coordinate curves are integral curves of $ \hat{L}_{z} $,
which can be easily seen from eq.~(\ref{eq:Killing-3:S2}).
Note that $\hat{L}_{z}$ spans a Cartan subalgebra of
$ \mathfrak{so}(3, \mathbb{R}) $, and that each $Y_{lm}$ spans a weight space
with respect to this Cartan subalgebra.

The above approach is very instructive.  Since the de~Sitter spacetime and
anti-de~Sitter spacetime are maximally symmetric, such an approach can be easily
applied to field equations in de~Sitter or anti-de~Sitter backgrounds.

In this paper we only take the Klein-Gordon equation on $ dS^{4} $,
the 4-dimensional de~Sitter spacetime (as a background), as an example.
The main idea can be illustrated as follows.

The symmetry group of $ dS^{4} $ is $ O(1, 4) $, whose Lie algebra
$ \mathfrak{so}(1, 4) $ induces a Lie algebra of Killing vector fields
on $ dS^{4} $.
Via these Killing vector fields, $ \mathfrak{so}(1, 4) $ acts on smooth
functions on $ dS^{4} $.  Thus $ C^{\infty}(dS^{4}) $, the space of all smooth
functions on $ dS^{4} $, is an $ \mathfrak{so}(1, 4) $-module.  (Equivalently
speaking, $ C^{\infty}(dS^{4}) $ is a representation space of
$ \mathfrak{so}(1, 4) $.)
Since the Klein-Gordon equation on $ dS^{4} $ is $ O(1, 4) $-invariant,
its solution space $ \mathscr{S}_{\textrm{KG}}(dS^{4}) $ is
$ O(1, 4) $-invariant.  Hence $ \mathscr{S}_{\textrm{KG}}(dS^{4}) $ is
an $ \mathfrak{so}(1, 4) $-submodule of $ C^{\infty}(dS^{4}) $.
It must be the direct sum of some irreducible $\mathfrak{so}(1, 4)$-submodules.
Therefore, to obtain the solution space of the Klein-Gordon equation
on $ dS^{4} $, we must construct these irreducible
$ \mathfrak{so}(1, 4) $-submodules of $ C^{\infty}(dS^{4}) $.

There is an important fact: the Klein-Gordon equation on $ dS^{4} $
is, in fact, an eigenvalue equation of the Casimir element of
$ \mathfrak{so}(1, 4) $.  See, eq.~(\ref{eq:KG:Casimir}) in this paper.
By aid of this fact, it is very easy to obtain all smooth solutions of
the Klein-Gordon equation in the de~Sitter background.  An interesting and
important consequence is that the mass in the Klein-Gordon equation is
discrete and imaginary, as shown in eq.~(\ref{eq:mass-spectrum}).

\section{Symmetry Group and Solution Space of the Klein-Gordon Equation
  on the de~Sitter Spacetime}
\label{sect:prog}

\subsection{Symmetry Group of the de~Sitter Spacetime}
\label{sect:symm-grp:dS}

Now we consider the 5-dimensional Minkowski space $\mathbb{R}^{1,4}$, with
$\xi^A$ ($A = 0$, $1$, \ldots, $4$) the Minkowski coordinates on it.
The de~Sitter spacetime of radius $l > 0$ can be treated as the hypersurface
\begin{equation}
  \eta_{AB} \, \xi^A \xi^B = - l^2
  \label{eq:dS}
\end{equation}
of $\mathbb{R}^{1, 4}$, where
$(\eta_{AB})_{5 \times 5} = \diag(1, -1, \dots, -1)$.
The linear group $O(1, 4)$ is the symmetry group of $\mathbb{R}^{1, 4}$,
leaving both the line element $\od s^2 = \eta_{AB} \, \od \xi^A \, \od \xi^B$
of $\mathbb{R}^{1, 4}$ and the hypersurface (\ref{eq:dS}) invariant.
Consequently, $O(1, 4)$ is also the symmetry group of $dS^4$.

For later usage, we describe the symmetries of $dS^4$ in some details.
The metric tensor field
$\widetilde{\vect{\eta}} := \eta_{AB} \, \od \xi^A \otimes \od \xi^B$
on $\mathbb{R}^{1,4}$ induces a metric tensor field $\tensor{g}$ on $dS^4$.
In fact, let $i \colon dS^4 \hookrightarrow \mathbb{R}^{1,4}$ be the inclusion,
then $\tensor{g} = i^* \widetilde{\vect{\eta}}$ is the pullback of
$\tilde{\vect{\eta}}$.  A linear transformation $D \in O(1, 4)$
on $\mathbb{R}^{1,4}$ leaves $dS^4$, the hypersurface (\ref{eq:dS}), invariant.
Thus we may set the restriction of $D$ to $dS^4$,
$\psi_D = D \at{dS^4} \colon dS^4 \rightarrow dS^4$,
to be a transformation on $ dS^{4} $.
It follows that $\psi_D$ is a symmetry of $(dS^4, \tensor{g})$.  That is,
$\psi_D$ is a diffeomorphism, satisfying
\begin{equation}
  \psi_D^* \tensor{g} = \tensor{g}
  \, .
  \label{eq:g-inv}
\end{equation}
Obviously, all $\psi_D$ with $D \in O(1, 4)$ form a group, which is isomorphic
to $O(1, 4)$.

In the following we describe the Lie algebras of the symmetry group.

First, there are the $5 \times 5$ matrices $X_{AB} \in \mathfrak{so}(1, 4)$
with $A$, $B = 0$, $1$, \ldots, $4$, whose $(C,D)^{\textrm{th}}$ entry
reads
\begin{equation}
  (X_{AB})^C_D = \delta^C_A \, \eta_{BD} - \delta^C_B \, \eta_{AD}
  \, .
  \label{def:X:AB}
\end{equation}
They satisfy $X_{AB} = - X_{BA}$ and
\begin{equation}
  [X_{AB}, X_{CD}]
  = \eta_{BC} \, X_{AD} + \eta_{AD} \, X_{BC}
  - \eta_{AC} \, X_{BD} - \eta_{BD} \, X_{AC}
  \, .
  \label{eq:comm:o(1,4)}
\end{equation}
Furthermore, all $X_{AB}$ with $A < B$ form a basis of $\mathfrak{so}(1, 4)$.

According to the theory of Lie groups, through the action of $O(1, 4)$
on $\mathbb{R}^{1,4}$, each matrix $X = (X^A_B) \in \mathfrak{so}(1, 4)$
generates a vector field
\begin{equation}
  \widetilde{\tensor{X}} := - X^A_B \, \xi^B \, \frac{\partial}{\partial \xi^A}
  \label{def:vec-field}
\end{equation}
on $\mathbb{R}^{1,4}$.
Let $\mathfrak{X}(\mathbb{R}^{1, 4})$ be the Lie algebra of smooth vector fields
on $\mathbb{R}^{1, 4}$.  Then the map
$ \mathfrak{so}(1, 4) \rightarrow \mathfrak{X}(\mathbb{R}^{1,4})$,
$ X \mapsto \tilde{\vect{X}} $ is a homomorphism of Lie algebras.
Roughly speaking, the above $\widetilde{\tensor{X}}$ is generated
by the infinitesimal transformation corresponding to the matrix $-X$.
In other words, the 1-parameter group of $\tilde{\tensor{X}}$ equals to
$\exp(- tX) \in O(1, 4)$.
So $\widetilde{\tensor{X}}$ is a Killing vector field on $\mathbb{R}^{1,4}$,
and vice versa, a Killing vector field on $\mathbb{R}^{1,4}$ corresponds to
a matrix $X \in \mathfrak{so}(1, 4)$ via eq.~(\ref{def:vec-field}).
For convenience, the Lie algebra consisting of all Killing vector fields on
$\mathbb{R}^{1,4}$ will be denoted by $\mathfrak{K}(\mathbb{R}^{1,4})$.
Then the map described in the above results in an isomorphism of Lie algebras
from $\mathfrak{so}(1, 4)$ to $\mathfrak{K}(\mathbb{R}^{1,4})$, mapping $[X, Y]$
for each pair of $X$, $Y \in \mathfrak{so}(1, 4)$ to the commutator
$[\widetilde{\tensor{X}}, \widetilde{\tensor{Y}}]$
of the corresponding Killing vector fields $\widetilde{\tensor{X}}$ and
$\widetilde{\tensor{Y}}$.
Especially, the matrix $X_{AB} \in \mathfrak{so}(1, 4)$, as defined
in eq.~(\ref{def:X:AB}), corresponds to the Killing vector field
\begin{equation}
  \widetilde{\tensor{X}}_{AB} = \xi_A \, \frac{\partial}{\partial \xi^B}
  - \xi_B \, \frac{\partial}{\partial \xi^A}
  \, ,
  \label{eq:XAB}
\end{equation}
where $\xi_A := \eta_{AB} \, \xi^B$.  Thus there are the commutators
\begin{equation}
  [\widetilde{\tensor{X}}_{AB}, \widetilde{\tensor{X}}_{CD}]
  = \eta_{BC} \, \widetilde{\tensor{X}}_{AD}
  + \eta_{AD} \, \widetilde{\tensor{X}}_{BC}
  - \eta_{AC} \, \widetilde{\tensor{X}}_{BD}
  - \eta_{BD} \, \widetilde{\tensor{X}}_{AC}
  \, .
\end{equation}
In fact, the above equations can be directly verified by virtue of
eq.~(\ref{eq:XAB}).

For an $ X \in \mathfrak{so}(1, 4) $, the corresponding Killing vector field
$\widetilde{\tensor{X}} \in \mathfrak{K}(\mathbb{R}^{1,4})$ is tangent to $dS^4$
at any point $\xi \in dS^4$.  Therefore such a vector field induces
a vector field on $dS^4$, denoted by $\tensor{X}$.
Roughly speaking, $\tensor{X}$ is the restriction of $\tilde{\tensor{X}}$
(as a section of the tangent bundle of $\mathbb{R}^{1, 4}$) to $dS^{4}$;
strictly speaking, $\tensor{X}$ is $i$-related\cite{Warner} to
$\tilde{\tensor{X}}$,
with $i \colon dS^{4} \hookrightarrow \mathbb{R}^{1, 4}$ being the inclusion.
In fact, if $X \in \mathfrak{so}(1, 4)$ corresponds to $\widetilde{\tensor{X}}$,
the 1-parameter group $\psi_{\exp(- t X)}$ is just generated by the vector field
$\tensor{X}$.
Hence, obviously, $\tensor{X}$ is a Killing vector field on $dS^4$.  It can be
verified that a Killing vector field on $dS^4$ corresponds to a matrix
in $\mathfrak{so}(1, 4)$.  
In this way, there exists the isomorphism of Lie algebras from
$\mathfrak{so}(1, 4)$ to $\mathfrak{K}(dS^4)$, where $\mathfrak{K}(dS^4)$
consists of all Killing vector fields on $dS^4$.  Especially, $X_{AB}$, hence
$\widetilde{\tensor{X}}_{AB}$, corresponds to a Killing vector field
$\tensor{X}_{AB}$ on $dS^4$, satisfying
\begin{equation}
  [\tensor{X}_{AB}, \tensor{X}_{CD}]
  = \eta_{BC} \, \tensor{X}_{AD} + \eta_{AD} \, \tensor{X}_{BC}
  - \eta_{AC} \, \tensor{X}_{BD} - \eta_{BD} \, \tensor{X}_{AC}
  \, .
\end{equation}

\subsection{Symmetries of the Klein-Gordon Equation on $dS^4$}

The Klein-Gordon equation on $dS^4$ is as shown in eq.~(\ref{eq:KG}).
Given a local coordinate system $(x^\mu)$ in $dS^4$, the expression of
$\tensor{X}_{AB}$ might be quite complicated.  Interestingly however, it can be
proved that, for an arbitrary smooth function $\phi$ on $dS^4$,
\begin{equation}
  g^{ab} \, \nabla_{a} \nabla_{b} \phi
  = - \frac{1}{2 l^2} \, \eta^{AC} \eta^{BD} \,
    L_{\tensor{X}_{AB}} L_{\tensor{X}_{CD}} \phi
  \, ,
\end{equation}
where $L_{\tensor{X}}$ is the Lie derivative with respect to a vector field
$\tensor{X}$.  Therefore, the Klein-Gordon equation (\ref{eq:KG}) is, indeed,
an eigenvalue equation
\begin{equation}
  C \phi = \frac{m^2 c^2 l^2}{\hbar^2}\, \phi
  \label{eq:KG:Casimir}
\end{equation}
for the Casimir operator
\begin{equation}
  C := \frac{1}{2} \, \eta^{AC} \eta^{BD} \,
    L_{\tensor{X}_{AB}} \, L_{\tensor{X}_{CD}}
\end{equation}
of the second order.\footnote{
A similar case on $AdS^{n + 1}$ can be found in \cite{ChangGuo}.
}

As mentioned in \S\ref{sect:symm-grp:dS}, an orthogonal transformation $D$
on $\mathbb{R}^{1,4}$ induces the automorphism $\psi_D$ of $(dS^4, \tensor{g})$,
whose pullback further induces a transformation
$\psi_D^* \colon C^{\infty}(dS^{4}) \rightarrow C^{\infty}(dS^{4})$,
mapping an arbitrary smooth function $\phi$ on $dS^4$ to another smooth function
$\phi' = \psi_D^* \phi := \phi \circ \psi_D$.
That is, at arbitrary point $\xi \in dS^4$,
\begin{equation}
  \phi'(\xi) = \phi(\psi_D (\xi)) = \phi(D \xi)
  \, .
\end{equation}
The group homomorphism mapping $D \in O(1, 4)$ to
$\psi_{D^{-1}}^* \in \mathrm{GL}(C^{\infty}(dS^{4}))$ is a
representation of $O(1, 4)$ on the vector space $C^{\infty}(dS^{4})$.
In other words, we have an action of $O(1, 4)$ on $ C^{\infty}(dS^{4}) $
on the left as follows:
$ O(1, 4) \times C^{\infty}(dS^{4}) \rightarrow C^{\infty}(dS^{4}) $,
$ (D, \phi) \mapsto \psi_{D^{-1}}^{*} \phi $.

Since the Klein-Gordon equation (\ref{eq:KG}) is determined by the metric tensor
field $\tensor{g}$, while $\psi_D^*$ leaves $\tensor{g}$ invariant
(see, eq.~(\ref{eq:g-inv})), it follows that, if $\phi$ is a smooth solution
of eq.~(\ref{eq:KG}), so is $\psi_D^* \phi$.
Let $\mathscr{S}_{\textrm{KG}}(dS^4)$ be the solution space of
eq.~(\ref{eq:dS}), namey, the set consisting of all smooth solutions of
eq.~(\ref{eq:KG}).  Then $\mathscr{S}_{\textrm{KG}}(dS^4)$ is a
vector space over $\mathbb{R}$ (for real-valued functions) or $\mathbb{C}$
(for complex-valued functions), being invariant under $ \psi_{D}^{*} $
for arbitrary $ D \in O(1, 4) $.  In other words, the action of $ O(1, 4) $
on $ C^{\infty}(dS^{4}) $ can be restricted to be
$ O(1, 4) \times \mathscr{S}_{\textrm{KG}}(dS^{4}) \rightarrow
  \mathscr{S}_{\textrm{KG}}(dS^{4})$,
$ (D, \phi) \mapsto \psi_{D^{-1}}^{*} \phi $.

\subsection{Smooth Solutions of the Klein-Gordon Equation on $dS^4$}
\label{sect:solution-space}

For any smooth function $ \phi $ on $ dS^{4} $ and any
$ X \in \mathfrak{so}(1, 4) $, the Lie derivative $L_{\tensor{X}} \phi$ of
$\phi$ with respect to $ \tensor{X} $ can be defined point-wisely~\cite{Warner}
by the derivative of $\psi_{\exp(- \lambda X)}^* \phi$ with respect to
the parameter $\lambda$, where the relation of $X$ and $\tensor{X}$ is
as described in \S\ref{sect:symm-grp:dS}.  For any $X \in \mathfrak{so}(1, 4)$
and any $\phi \in C^{\infty}(dS^4)$, the action of $ X $ upon $ \phi $
results in $ X.\phi $, defined by
\begin{equation}
  X.\phi := L_{\tensor{X}} \phi = \tensor{X} \phi
  \, .
  \label{action:X.phi}
\end{equation}
Obviously, $ X.\phi $ is still a smooth function on $ dS^{4} $.
Hence $ C^{\infty}(dS^{4}) $ becomes an $ \mathfrak{so}(1, 4) $-module.

Especially, if $ \phi $ is a smooth solution of the Klein-Gordon equation
(\ref{eq:KG}), so is $ X.\phi $ for any $ X \in \mathfrak{so}(1, 4) $.
Thus $ \mathscr{S}_{\textrm{KG}}(dS^{4}) $ is an $\mathfrak{so}(1, 4)$-submodule
of $ C^{\infty}(dS^{4}) $.

In \S\ref{sect:Killing-form} we shall show that $\mathfrak{so}(1, 4)$ is
semisimple.
According to the representation theory of Lie algebras~\cite{Humphreys},
the solution space $\mathscr{S}_{\textrm{KG}}(dS^4)$ can be decomposed into
the direct sum of irreducible $\mathfrak{so}(1, 4)$-submodules.  Hence every
solution $\phi$ of eq.~(\ref{eq:KG}) can be decomposed into the sum of
finite many functions $\phi_1$, \ldots, $\phi_k$, with $\phi_i$ ($i = 1$,
\ldots, $k$) belonging to certain an irreducible
$\mathfrak{so}(1, 4)$-submodule.

Since eq.~(\ref{eq:KG:Casimir}) is equivalent to eq.~(\ref{eq:KG}) and
\begin{equation}
  [C, L_{\tensor{X}}] = 0
  \, , \qquad
  \forall \tensor{X} \in \mathfrak{K}(dS^4)
  \, ,
\end{equation}
it follows Schur's lemma~\cite{Humphreys} that the solution space
$\mathscr{S}_{\textrm{KG}}(dS^{4})$ is the direct sum of irreducible
$\mathfrak{so}(1, 4)$-submodules belonging to the same eigenvalue of $C$.
In other words, any function $\phi$ belonging to an irreducible
$\mathfrak{so}(1, 4)$-submodule of $C^\infty(dS^4)$ is a smooth solution of
the Klein-Gordon equation (\ref{eq:KG:Casimir}) with certain a mass.

Therefore, in order to find smooth solutions of the Klein-Gordon equation
on $dS^4$, it is necessary to find the irreducible
$\mathfrak{so}(1, 4)$-submodule of $C^\infty(dS^4)$.

A consequence of the above conclusion is that the mass of the Klein-Gordon
equation cannot be arbitrary.  The mass must be related to an eigenvalue of
$ C $, which in turn is determined by the highest weights of irreducible
$ \mathfrak{so}(1, 4) $-submodules contained in
$ \mathscr{S}_{\textrm{KG}}(dS^{4}) $.
In \S\ref{sect:solution-mass} we can calculate the mass, as shown in
eq.~(\ref{eq:mass-spectrum}), where the nonnegative integer $N$ specifies
the highest weight of an irreducible $ \mathfrak{so}(1, 4) $-submodule.
Furthermore, in \S\ref{sect:solution-mass} we shall prove that different
highest weight (or equivalently, $ N $) corresponds to different mass.
It follows that the solution space $ \mathscr{S}_{\textrm{KG}}(dS^{4}) $
is nothing but an irreducible $ \mathfrak{so}(1, 4) $-submodule of
$ C^{\infty}(dS^{4}) $.

\section{Structure of the Lie Algebra $\mathfrak{so}(1, 4)$}
\label{sect:Lie-alg}

There are papers discussing irreducible representations of
$\mathfrak{so}(1, 4)$, such as \cite{LHThomas} and \cite{TDNewton}.
In \cite{LHThomas}, unitary representations of $\mathfrak{so}(1, 4)$
were constructed out of irreducible representations of
$\mathfrak{so}(4, \mathbb{R})$, similar to the method by Wigner\cite{Wigner}.
The resulted irreducible unitary representations are infinite
dimensional\cite{TDNewton}.  Since we are seeking the solution space of
the Klein-Gordon equation, in this paper we are not interested in
the representation of $\mathfrak{so}(1, 4)$, but rather its representation
spaces, irreducible $\mathfrak{so}(1, 4)$-submodules in $C^{\infty}(dS^{4})$.
For this purpose, the framework in \cite{LHThomas} is so complicated
in practice.

In this paper we construct irreducible $\mathfrak{so}(1, 4)$-submodules of
$C^{\infty}(dS^{4})$ using the standard methods in the theory of
Lie groups\cite{Warner} and Lie algebras\cite{Humphreys}.

In this section we briefly describe the abstract structure of
$\mathfrak{so}(1, 4)$ and/or its complexification,
$\mathfrak{L} = \mathfrak{so}(1, 4) \otimes_{\mathbb{R}} \mathbb{C}$.
In the next section we apply the representation theory to
the $\mathfrak{so}(1, 4)$-module $C^{\infty}(dS^{4})$.

\subsection{The Killing Form of $\mathfrak{so}(1, 4)$}
\label{sect:Killing-form}

In this paper, the Killing form of $\mathfrak{so}(1, 4)$ is denoted by $\kappa$.
It is easy to verify that\footnote{
For $\mathfrak{so}(p, q)$ with integers $p \geqslant 0$ and $q \geqslant 0$,
the Lie brackets also satisfy eq.~(\ref{eq:comm:o(1,4)}).
By virtue of eq.~(\ref{eq:comm:o(1,4)}) and the definition of the Killing form,
one can obtain the formula
\begin{equation}
  \kappa(X_{AB}, X_{CD})
  = 2(n - 2) \, (\eta_{AD} \, \eta_{BC} - \eta_{AC} \, \eta_{BD})
\end{equation}
for $\mathfrak{so}(p, q)$, where $n = p + q \geqslant 2$.
}
\begin{equation}
  \kappa(X_{AB}, X_{CD})
  = 6 \, (\eta_{AD} \, \eta_{BC} - \eta_{AC} \, \eta_{BD})
  \, .
\end{equation}
This tells us that (1)~for $X_{AB}$ with $A < B$, they are mutually
orthogonal (with respect to the Killing form), and that (2)~the Killing form of
$X_{AB}$ with itself is $\mp 6$.  A corollary is that the Killing form is
nondegenerate.  Hence $\mathfrak{so}(1, 4)$ is a semisimple Lie algebra.

\subsection{The Abstract Root System of
  $\mathfrak{L} = \mathfrak{so}(1, 4) \otimes_{\mathbb{R}} \mathbb{C}$}

According to the theory of Lie algebras, if a semisimple Lie algebra
$\mathfrak{L}$ is over an algebraically closed field of characteristic 0,
such as $\mathbb{C}$, the abstract root system $\Phi$ of $\mathfrak{L}$
is determined by $\mathfrak{L}$ itself.  See, for example, \mbox{\S 16} in
\cite{Humphreys}.  However, since the Lie algebra $\mathfrak{so}(1, 4)$ is
over $\mathbb{R}$, which is not an algebraically closed field, the theory of
root systems for semisimple Lie algebras cannot be applied to it.
Therefore we use its complexification
$\mathfrak{L} := \mathfrak{so}(1, 4) \otimes_{\mathbb{R}} \mathbb{C}$ instead.

All the following results can be obtained using the standard approach
in the theory of Lie algebra\cite{Humphreys}.  We neglect all these processes,
listing the results only.

\begin{figure}[hbt]
  \centering
  \includegraphics{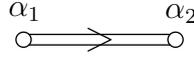}
  \caption{The Dynkin diagram of
    $\mathfrak{L} = \mathfrak{so}(1, 4) \otimes_{\mathbb{R}} \mathbb{C}$.}
  \label{fig:Dynkin}
\end{figure}

The Dynkin diagram of $\mathfrak{L}$ is as shown in Figure~\ref{fig:Dynkin}.
Therefore, $\mathfrak{L}$ is a semisimple Lie algebra of type
$\mbox{\textsf{B}}_2$.  Let $\Delta = \{\alpha_1, \alpha_2 \}$ be the base of
the root system $\Phi$ of $\mathfrak{L}$.  Then its Cartan matrix reads
\begin{equation}
  \big( \langle \alpha_i, \alpha_j\rangle \big)_{2 \times 2}
  = \mathmatrix{rr}{
      2 & -2 \\
      -1 & 2
    }
  \, .
  \label{eq:Cartan-matrix}
\end{equation}
Here the Cartan integer related to two roots $\alpha$ and $\beta$ is
\begin{equation}
  \langle \alpha, \beta\rangle = \frac{2 \, (\alpha, \beta)}{(\beta, \beta)}
  \, ,
\end{equation}
where $(\cdot, \cdot)$ is the inner product on the Euclidean space spanned by
all the roots of $\mathfrak{L}$.

The root system $\Phi$ of $\mathfrak{L}$ consists of the following roots:
$\pm \alpha_1$, $\pm \alpha_2$, $\pm (\alpha_1 + \alpha_2)$ and
$\pm (\alpha_1 + 2 \, \alpha_2)$.
For later reference the set of positive roots is denoted by $ \Phi^{+} $,
reading
\begin{equation}
  \Phi^+ = \{ \alpha_1, \ \alpha_2, \alpha_1 + \alpha_2,
    \ \alpha_1 + 2 \, \alpha_2 \}
  \, .
\end{equation}
Then $\Phi = \Phi^+ \cup \Phi^-$, where $\Phi^- = - \Phi^+$ is the set of
negative roots.

\begin{figure}[hbt]
  \centering
  \includegraphics[width = 70mm, height = 60mm]{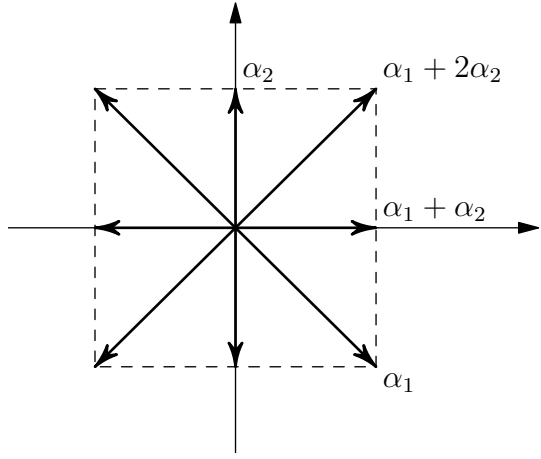}
  \caption{The root system of
    $\mathfrak{L} = \mathfrak{so}(1, 4) \otimes_{\mathbb{R}} \mathbb{C}$.}
  \label{fig:roots}
\end{figure}

Let $E_{\Phi}$ be the 2-dimensional Euclidean space spanned by the root system
$\Phi$.  Obviously, $\alpha_1$ and $\alpha_2$ form a basis of $E_{\Phi}$.
In this paper, the inner product on $ E_{\Phi} $ is resulted in
from the Killing form $ \kappa $, as follows:
We first choose a Cartan subalgebra $ \mathfrak{h} $ of $ \mathfrak{L} $.
Then the restriction of $ \kappa $ to $ \mathfrak{h} $ is
nondegenerate~\cite{Humphreys}, inducing a nondegenerate bilinear form
$ (\cdot, \cdot) $ on $E_{\mathfrak{h}^*}$, satisfying
\begin{equation}
  (\alpha_1, \alpha_1) = \frac{1}{3} \, , \qquad
  (\alpha_1, \alpha_2) = - \frac{1}{6} \, , \qquad
  (\alpha_2, \alpha_2) = \frac{1}{6} \, .
  \label{eq:inner-product}
\end{equation}
Then such a bilinear form is turned to be the inner product on $ E_{\Phi} $.
With the aid of the data in eqs.~(\ref{eq:inner-product}), the roots can be
drawn as in Figure~\ref{fig:roots}.

The Cartan subalgebra (also a maximal toral subalgebra) of a semisimple
Lie algebra is not unique.  Being conjugate to each other~\cite{Humphreys},
none of these Cartan subalgebras is more significant than others.
However, when the action of the Lie algebra on a differential manifold is
considered, this is no longer the same situation.  So, for
$\mathfrak{L} = \mathfrak{so}(1, 4) \otimes_{\mathbb{R}} \mathbb{C}$,
we shall take two typical Cartan subalgebras into account.

\subsection{The Cartan Subalgebra and Root Spaces (I)}
\label{sect:h}

The first Cartan subalgebra is spanned by $X_{12}$ and $X_{04}$, and denoted by
$\mathfrak{h}$ in this paper.  With respect to $\mathfrak{h}$, the roots
$\alpha_1$ and $\alpha_2 \in \mathfrak{h}^*$ satisfy
\begin{alignat}{2}
  \alpha_1(X_{12}) &= i, \qquad &
  \alpha_1(X_{04}) &= -1,
  \label{eq:alpha1:h}
  \\
  \alpha_2(X_{12}) &= 0,  &
  \alpha_2(X_{04}) &= 1.
  \label{eq:alpha2:h}
\end{alignat}

For each positive root $\beta \in \Phi^+$, the root spaces of $\beta$ and
$-\beta$ are denoted by $ \mathfrak{L}_{\pm\beta} $, respectively.
The bases of $\mathfrak{L}_{\pm\beta}$ are denoted by $ e_{\beta} $ and
$ f_{\beta} $, respectively.  They are chosen as follows:
\begin{alignat}{2}
  e_{\alpha_1} &= \frac{1}{2} \, (X_{01} - i \, X_{02} - X_{14} + i \, X_{24})
  \, , \qquad &
  f_{\alpha_1} &= \frac{1}{2} \, (X_{01} + i \, X_{02} + X_{14} + i \, X_{24})
  \, ,
  \label{eq:Chevalley-1:I}
  \\
  e_{\alpha_2} &= X_{03} + X_{34}
  \, , &
  f_{\alpha_2} &= X_{03} - X_{34}
  \, ,
  \\
  e_{\alpha_1 + \alpha_2} &= - X_{13} + i \, X_{23}
  \, , &
  f_{\alpha_1 + \alpha_2} &= X_{13} + i \, X_{23}
  \, ,
  \\
  e_{\alpha_1 + 2 \alpha_2}
  &= - \frac{1}{2} \, (X_{01} - i \, X_{02} + X_{14} - i \, X_{24})
  \, , \quad &
  f_{\alpha_1 + 2 \alpha_2}
  &= - \frac{1}{2} \, (X_{01} + i \, X_{02} - X_{14} - i \, X_{24})
  \, .
  \label{eq:Chevalley-4:I}
\end{alignat}
It can be verified that these generators, together with
\begin{equation}
  h_{\alpha_1} = - X_{04} - i \, X_{12}
  \, , \qquad
  h_{\alpha_2} = 2 \, X_{04}
  \, ,
  \label{eq:h:I}
\end{equation}
form a Chevalley basis of $\mathfrak{L}$.  For the commutators between them,
see Appendix~\ref{sect:commutators}.

\subsection{The Cartan Subalgebra and Root Spaces (II)}
\label{sect:h'}

The second typical Cartan subalgebra of $\mathfrak{L}$ is spanned by $X_{12}$
and $X_{34}$, and denoted by $\mathfrak{h}'$ in this paper.  The roots
$\alpha_1$ and $\alpha_2 \in \mathfrak{h}'^*$ satisfy
\begin{alignat}{2}
  \alpha_1(X_{12}) &= i
  \, , \qquad &
  \alpha_1(X_{34}) &= -i
  \, ,
  \label{eq:alpha1:h'}
  \\
  \alpha_2(X_{12}) &= 0
  \, , &
  \alpha_2(X_{34}) &= i
  \, .
  \label{eq:alpha2:h'}
\end{alignat}
A Chevalley basis of $\mathfrak{L}$ can be chosen as follows:
\begin{alignat}{2}
  h_{\alpha_1} &= -i \, X_{12} + i \, X_{34}
  \, , &
  h_{\alpha_2} &= - 2 i \, X_{34}
  \, ,
  \label{eq:Chevalley-1:II}
  \\
  e_{\alpha_1} &= \frac{1}{2} \, (X_{13} + i \, X_{14} - i \, X_{23} + X_{24})
  \, , &
  f_{\alpha_1} &= - \frac{1}{2} \, (X_{13} - i \, X_{14} + i \, X_{23} + X_{24})
  \, , \\
  e_{\alpha_2} &= X_{03} - i \, X_{04}
  \, , &
  f_{\alpha_2} &= X_{03} + i \, X_{04}
  \, , \\
  e_{\alpha_1 + \alpha_2} &= - X_{01} + i \, X_{02}
  \, , &
  f_{\alpha_1 + \alpha_2} &= - X_{01} - i \, X_{02}
  \, , \\
  e_{\alpha_1 + 2 \alpha_2}
  &= - \frac{1}{2} \, (X_{13} - i \, X_{14} - i \, X_{23} - X_{24})
  \, , \qquad &
  f_{\alpha_1 + 2 \alpha_2}
  &= \frac{1}{2} \, (X_{13} + i \, X_{14} + i \, X_{23} - X_{24})
  \, .
  \label{eq:Chevalley-5:II}
\end{alignat}
Their commutators are shown as in Appendix~\ref{sect:commutators}.

\section{Irreducible $\mathfrak{L}$-Modules of Smooth Functions}
\label{sect:irred-modules}

\subsection{Local Coordinates Adapted to $\mathfrak{h}$ or $\mathfrak{h}'$}
\label{sect:coord}

For a given PDE, it is not that every coordinate system is suitable
for separation of variables.  A suitable coordinate system must be adapted to
the symmetry group of the PDE.

In this subsection we try to find suitable coordinate systems on $dS^4$ that is
adapted to the symmetry group $O(1, 4)$.  Such a coordinate system is based on
the congruence of integral submanifolds of a Cartan subalgebra.
Since two typical Cartan subalgebras are presented in this paper,
there are different coordinate systems corresponding to these
Cartan subalgebras, respectively.

The first type of coordinates are lated to the Cartan subalgebra $\mathfrak{h}$
in \S\ref{sect:h}.
For both vector fields $\widetilde{\tensor{X}}_{04}$ and
$\widetilde{\tensor{X}}_{12}$, their integral curves in $\mathbb{R}^{1,4}$
can be described by
\begin{alignat}{1}
  \xi^0 & = T \cosh \chi + X \sinh \chi
  \, ,
  \label{eq:xi0}
  \\
  \xi^4 & = T \sinh \chi + X \cosh \chi
  \, , \\
  \xi^1 & = Y \cos \varphi + Z \sin \varphi
  \, , \\
  \xi^2 & = - Y \sin \varphi + Z \cos \varphi
  \, , \\
  \xi^3 & = \Xi
  \, .
  \label{eq:xi3}
\end{alignat}
For a set of fixed $ T $, $ X $, $ Y $, $ Z $, $ \Xi $ and
$ \varphi \in \mathbb{R} $,
the above equations are parameter equations of an integral curve of
$\widetilde{\tensor{X}}_{04}$, with $\chi$ the curve parameter;
for a set of fixed $T$, $X$, $Y$, $Z$, $\Xi$ and $\chi \in \mathbb{R}$,
they are the parameter equations of an integral curve of
$\widetilde{\tensor{X}}_{12}$, with $\varphi$ the curve parameter.

For fixed $ T $, $ X $, $ Y $, $ Z $ and $ \Xi \in \mathbb{R} $, when both
$ \chi $ and $ \varphi $ are viewed as parameters, the above equations describe
a 2-surface in $ \mathbb{R}^{1, 4} $.  Note that, when $ T = X = 0 $ or
$ Y  = Z = 0 $, such a 2-surface may be degenerate into a curve or even a point.
All these surfaces, no matter nondegenerate or not, can be viewed as integral
submanifolds of the Cartan subalgebra $ \mathfrak{h} $.
Obviously, such an integral submanifold is contained in $dS^4$ if and only if
\begin{equation}
  T^2 - X^2 - Y^2 - Z^2 - \Xi^2 = - l^2
  \, .
  \label{condition:int-submfd:dS}
\end{equation}
For a region of $dS^4$ where both $\widetilde{\tensor{X}}_{04}$ and
$\widetilde{\tensor{X}}_{12}$ are nonzero everywhere, the parameter $\chi$ and
$\varphi$ of the integral submanifolds can be developed into a local coordinate
system $(\chi, \zeta, \theta, \varphi)$ of $dS^4$ by setting
$T = T(\zeta, \theta)$, $X = X(\zeta, \theta)$, and so on,
in eq.~(\ref{condition:int-submfd:dS}).
Since $T$, $X$, $Y$ and $Z$ are redundant, some of them can be set to be zero
directly.
For example, in the region $|\xi^0| > |\xi^4|$ of $dS^4$, it follows that
$(\xi^1)^2 + (\xi^2)^2 + (\xi^3)^3 > l^2$.  We may choose certain functions
$T(\zeta, \theta)$, $X(\zeta, \theta)$, etc., so that
eqs.~(\ref{eq:xi0}) to (\ref{eq:xi3}) turn out to be
\begin{alignat}{1}
  \xi^0 & = l \sinh\zeta \cosh\chi
  \, ,
  \label{eq:xi0:U1}
  \\
  \xi^1 & = l \cosh\zeta \sin\theta \cos\varphi
  \, , \\
  \xi^2 & = l \cosh\zeta \sin\theta \sin\varphi
  \, , \\
  \xi^3 & = l \cosh\zeta \cos\theta
  \, , \\
  \xi^4 & = l \sinh\zeta \sinh\chi
  \, .
  \label{eq:xi4:U1}
\end{alignat}
For another example, on the region $|\xi^4| > |\xi^0|$ of $dS^4$, the local
coordinate system $(\chi, \zeta, \theta, \varphi)$ can be such that
\begin{alignat}{1}
  \xi^0 & = l \cos\zeta \sinh\chi
  \, ,
  \label{eq:xi0:U2}
  \\
  \xi^1 & = l \sin\zeta \sin\theta \cos\varphi
  \, , \\
  \xi^2 & = l \sin\zeta \sin\theta \sin\varphi
  \, , \\
  \xi^3 & = l \sin\zeta \cos\theta
  \, , \\
  \xi^4 & = l \cos\zeta \cosh\chi
  \, .
  \label{eq:xi4:U2}
\end{alignat}

The above coordinate systems are adapted to the Cartan subalgebra
$\mathfrak{h} = \mathspan_{\mathbb{C}} \{ X_{04}, X_{12} \}$.
In the similar way, we can find some coordinate systems adapted to
$\mathfrak{h}' = \mathspan_{\mathbb{C}} \{ X_{12}, X_{34} \}$.
For example, one such coordinate system $(\chi, \zeta, \theta, \varphi)$
is defined by
\begin{alignat}{1}
  \xi^0 &= l \sinh\chi
  \, ,
  \label{eq:coord-sys:U:0}
  \\
  \xi^1 &= l \cosh\chi \cos\zeta \cos\theta
  \, ,
  \label{eq:coord-sys:U:1}
  \\
  \xi^2 &= l \cosh\chi \cos\zeta \sin\theta
  \, ,
  \label{eq:coord-sys:U:2}
  \\
  \xi^3 &= l \cosh\chi \sin\zeta \cos\varphi
  \, , \\
  \xi^4 &= l \cosh\chi \sin\zeta \sin\varphi
  \, .
  \label{eq:coord-sys:U:4}
\end{alignat}
In such a coordinate system, the $\theta$-coordinate curves are integral curves
of $\tensor{X}_{12}$, and the $\varphi$-coordinate curves are those of
$\tensor{X}_{34}$.
Coordinate neighborhoods of $ (\chi, \zeta, \theta, \varphi) $ should be like
these: for $ j = 0 $, $ 1 $, $ 2 $ and $ 3 $, respectively,
$ U_{j00} $, where $ \frac{j}{2} \pi < \zeta < \frac{j + 1}{2} \pi $,
$ - \pi < \theta < \pi $ and $ - \pi < \varphi < \pi $;
$ U_{j01} $, where $ \frac{j}{2} \pi < \zeta < \frac{j + 1}{2} \pi $,
$ - \pi < \theta < \pi $ and $ 0 < \varphi < 2 \pi $;
$ U_{j10} $, where $ \frac{j}{2} \pi < \zeta < \frac{j + 1}{2} \pi $,
$ 0 < \theta < 2 \pi $ and $ - \pi < \varphi < \pi $;
$ U_{j11} $, where $ \frac{j}{2} \pi < \zeta < \frac{j + 1}{2} \pi $,
$ 0 < \theta < 2 \pi $ and $ 0 < \varphi < 2 \pi $.
The union
$ U =  \bigcup_{j = 0}^{3} (U_{j00} \cup U_{j01} \cup U_{j10} \cup U_{j11})$
is an open subset of $ dS^{4} $.  It is not connected, with
$ U_{j00} \cup U_{j01} \cup U_{j10} \cup U_{j11} $
(for each $ j = 0 $, $ 1 $, $ 2 $ and $ 3 $) a connected component.

\subsection{Verma Modules of Smooth Functions}
\label{sect:Verma-modules}

When $X_{AB} \in \mathfrak{so}(1, 4)$ is mapped to the Killing vector field
$ \tensor{X}_{AB} $ on $ dS^{4} $ by the isomorphism from $\mathfrak{so}(1, 4)$
to $ \mathfrak{K}(dS^{4}) $, (see, \S\ref{sect:symm-grp:dS}),
by linearity $h_\beta$, $e_\beta$ and $f_\beta$ for $\beta \in \Phi^+$ are
mapped to corresponding complex vector fields $\tensor{h}_\beta$,
$\tensor{e}_\beta$ and $\tensor{f}_\beta$ on $dS^4$, respectively.
For details, see eqs.~(\ref{eq:Chevalley-1:I}) to (\ref{eq:h:I})
and eqs.~(\ref{eq:Chevalley-1:II}) to (\ref{eq:Chevalley-5:II}).
These complex vector fields can be expressed in terms of the coordinates $\chi$,
$\zeta$, $\theta$ and $\varphi$.  For $ h_{\beta} $, $ e_{\beta} $ and
$ f_{\beta} $ with respect to the Cartan subalgebra $ \mathfrak{h}' $,
the coordinate expressions of corresponding $ \tensor{h}_{\beta} $,
$ \tensor{e}_{\beta} $ and $ \tensor{f}_{\beta} $ are, respectively,
\begin{alignat}{1}
  \tensor{h}_{\alpha_1}
  &= i \, \frac{\partial}{\partial\theta}
  - i \, \frac{\partial}{\partial\varphi}
  \, ,
  \label{eq:h1:U}
  \\
  \tensor{h}_{\alpha_2}
  &= 2i \, \frac{\partial}{\partial\varphi}
  \, ,
  \label{eq:h2:U}
  \\
  \tensor{e}_{\alpha_1}
  &= - \frac{e^{i \varphi - i \theta}}{2} \, \Big(
      \frac{\partial}{\partial\zeta}
      + i \, \tan\zeta \, \frac{\partial}{\partial\theta}
      + i \, \cot\zeta \, \frac{\partial}{\partial\varphi}
    \Big)
  \, ,
  \label{eq:e1:U}
  \\
  \tensor{f}_{\alpha_1}
  &= \frac{e^{i \theta -i \varphi}}{2} \, \Big(
      \frac{\partial}{\partial\zeta}
      - i \, \tan\zeta \, \frac{\partial}{\partial\theta}
      - i \, \cot\zeta \, \frac{\partial}{\partial\varphi}
    \Big)
  \, ,
  \label{eq:f1:U}
  \\
  \tensor{e}_{\alpha_2}
  &= e^{-i \varphi} \Big(
      \sin\zeta \, \frac{\partial}{\partial\chi}
      + \tanh\chi \cos\zeta \, \frac{\partial}{\partial\zeta}
      - i \, \frac{\tanh\chi}{\sin\zeta} \, \frac{\partial}{\partial\varphi}
    \Big)
  \, ,
  \label{eq:e2:U}
  \\
  \tensor{f}_{\alpha_2}
  &= e^{i \varphi} \Big(
      \sin\zeta \, \frac{\partial}{\partial\chi}
      + \tanh\chi \cos\zeta \, \frac{\partial}{\partial\zeta}
      + i \, \frac{\tanh\chi}{\sin\zeta} \, \frac{\partial}{\partial\varphi}
    \Big)
  \, ,
  \label{eq:f2:U}
  \\
  \tensor{e}_{\alpha_1 + \alpha_2}
  &= e^{-i \theta} \Big(
      - \cos\zeta \, \frac{\partial}{\partial\chi}
      + \tanh\chi \sin\zeta \, \frac{\partial}{\partial\zeta}
      + i \, \frac{\tanh\chi}{\cos\zeta} \, \frac{\partial}{\partial\theta}
    \Big)
  \, ,
  \label{eq:e3:U}
  \\
  \tensor{f}_{\alpha_1 + \alpha_2}
  &= e^{i \theta} \Big(
      - \cos\zeta \, \frac{\partial}{\partial\chi}
      + \tanh\chi \sin\zeta \, \frac{\partial}{\partial\zeta}
      - i \, \frac{\tanh\chi}{\cos\zeta} \, \frac{\partial}{\partial\theta}
    \Big)
  \, ,
  \label{eq:f3:U}
  \\
  \tensor{e}_{\alpha_1 + 2 \alpha_2}
  &= \frac{e^{-i \theta - i \varphi}}{2} \, \Big(
      \frac{\partial}{\partial\zeta}
      + i \, \tan\zeta \, \frac{\partial}{\partial\theta}
      - i \, \cot\zeta \, \frac{\partial}{\partial\varphi}
    \Big)
  \, ,
  \label{eq:e4:U}
  \\
  \tensor{f}_{\alpha_1 + 2 \alpha_2}
  &= - \frac{e^{i \theta + i \varphi}}{2} \, \Big(
      \frac{\partial}{\partial\zeta}
      - i \, \tan\zeta \, \frac{\partial}{\partial\theta}
      + i \, \cot\zeta \, \frac{\partial}{\partial\varphi}
    \Big)
  \, .
  \label{eq:f4:U}
\end{alignat}

In the sense of eq.~(\ref{action:X.phi}), real smooth functions on $dS^4$
form an $\mathfrak{so}(1,4)$-module $C^\infty (dS^4)$, and complex
smooth functions on $dS^4$ form an $\mathfrak{L}$-module
(where $ \mathfrak{L} = \mathfrak{so}(1, 4) \otimes_{\mathbb{R}} \mathbb{C} $),
denoted by $\mathscr{C} (dS^4)$.  Obviously,
$\mathscr{C} (dS^4) = C^\infty (dS^4) \otimes_{\mathbb{R}} \mathbb{C}$.

We are interested in finite dimensional irreducible $\mathfrak{L}$-submodules
of $\mathscr{C} (dS^4)$, because the solution space of the Klein-Gordon equation
is the direct sum of certain irreducible $\mathfrak{L}$-submodules (for complex
solutions) or $ \mathfrak{so}(1, 4) $-submodules (for real solutions).
See, \S\ref{sect:solution-space}.
These irreducible $\mathfrak{L}$-submodules can be obtained using the standard
method in the representation theory of Lie algebras~\cite{Humphreys}.

In this subsection we shall show these $\mathfrak{L}$-submodules related to
the Cartan subalgebra $\mathfrak{h}'$.

For the roots $\alpha_1$ and $\alpha_2 \in \mathfrak{h}'^*$ as shown in
eqs.~(\ref{eq:alpha1:h'}) and (\ref{eq:alpha2:h'}), let $\lambda_1$ and
$\lambda_2 \in \mathfrak{h}^*$ be the fundamental dominant weights, defined by
\begin{equation}
  \langle \lambda_i, \alpha_j \rangle
  = \lambda_i (h_{\alpha_j})
  = \delta_{ij}
  \, , \qquad
  (i, j = 1, 2) \, .
  \label{eq:fdmt-dom-weight}
\end{equation}
Then it follows that $(\lambda_1, \lambda_2)$ is the dual basis of
$(h_{\alpha_1}, h_{\alpha_2})$.
By virtue of the Cartan matrix (\ref{eq:Cartan-matrix}), it is easy to obtain
\begin{equation}
  \lambda_1 = \alpha_1 + \alpha_2
  \, , \qquad
  \lambda_2 = \frac{1}{2} \, \alpha_1 + \alpha_2
  \, .
  \label{eq:lambda-alpha}
\end{equation}

Let $n_1$ and $n_2$ be two integers.  Then the weight
$\mu = n_1 \lambda_1 + n_2 \lambda_2$ satisfies
\begin{equation}
  \mu (h_{\alpha_i}) = n_i
  \, , \qquad
  (i = 1, 2) \, .
  \label{eq:mu:h-i}
\end{equation}
A function $\phi_\mu \in \mathscr{C}(dS^4)$ of weight $\mu$ satisfies
\begin{equation}
  \tensor{h}_{\alpha_i} \phi_\mu
  = \mu(h_{\alpha_i}) \, \phi_\mu
  = n_i \, \phi_\mu
  \, , \qquad
  (i = 1, 2) \, .
\end{equation}
On account of the expressions (\ref{eq:h1:U}) and (\ref{eq:h2:U}), we have
\begin{displaymath}
  \phi_\mu = \Phi_\mu(\chi, \zeta) \,
    e^{- \frac{n_2}{2} \, i \, \varphi - (n_1 + \frac{n_2}{2}) i \, \theta}
  \, ,
\end{displaymath}
with $\Phi_\mu(\chi, \zeta)$ an unknown function.

Let $\lambda = N_1 \lambda_1 + N_2 \lambda_2$ be the highest weight.
Then
  $ \tensor{e}_{\alpha_1} \phi_\lambda = 0 $
and
  $ \tensor{e}_{\alpha_1 + 2 \alpha_2} \phi_\lambda = 0 $
result in, respectively,
\begin{alignat*}{1}
  \frac{\partial}{\partial \zeta} \Phi_\lambda(\chi, \zeta)
  + \Big( N_1 + \frac{N_2}{2} \Big) \, \Phi_\lambda(\chi, \zeta) \tan\zeta
  + \frac{N_2}{2} \, \Phi_\lambda(\chi, \zeta) \cot\zeta
  &= 0
  \, ,
  \\
  \frac{\partial}{\partial \zeta} \Phi_\lambda(\chi, \zeta)
  + \Big( N_1 + \frac{N_2}{2} \Big) \, \Phi_\lambda(\chi, \zeta) \tan\zeta
  - \frac{N_2}{2} \, \Phi_\lambda(\chi, \zeta) \cot\zeta
  &= 0
  \, .
\end{alignat*}
These equations imply that $N_2 = 0$, and that
\[
  \frac{\partial}{\partial \zeta} \Phi_\lambda(\chi, \zeta)
  + N_1 \Phi_\lambda(\chi, \zeta) \tan\zeta = 0
  \, .
\]
The general solution for this equation is
\[
  \Phi_\lambda(\chi, \zeta) = X(\chi)  \big( \cos\zeta \big)^{N_1}
  \, ,
\]
with $X(\chi)$ an unknown function of $ \chi $.
Furthermore, $\tensor{e}_{\alpha_2} \phi_\lambda = 0$ results in
\[
  X'_\lambda(\chi) - N_1 X_\lambda(\chi) \tanh\chi = 0
  \, .
\]
It can be checked that no more relations can be obtained from
$\tensor{e}_{\alpha_1 + \alpha_2} \phi_\lambda = 0$.
Hence
\[
  X_\lambda(\chi) = C \big( \cosh\chi \big)^{N_1}
  \, ,
\]
where $C$ is an integral constant.

{From} now on $N_1$ will be denoted by $N$.  Then the highest weight reads
\begin{equation}
  \lambda = N \lambda_1
  \, .
\end{equation}
Fixing the integral constant, we can select
\begin{equation}
  \phi_\lambda = \big( \cosh\chi \cos\zeta \, e^{-i \theta} \big)^{N}
  \label{eq:phi-lambda}
\end{equation}
as the function of the highest weight $\lambda$.

Recursively, we can verify that, for nonnegative integers $j$, $k$ and $l$,
\begin{alignat}{1}
  \phi_\lambda^{(jkl)} & := L_{\mathbf{f}_{\alpha_1 + \alpha_2}}^j
    L_{\mathbf{f}_{\alpha_1 + 2 \alpha_2}}^k
    L_{\mathbf{f}_{\alpha_1}}^l \phi_\lambda
  \nonumber \\
  &= \sum_{j' = 0}^{\lfloor\frac{j}{2}\rfloor} \sum_{k' = 0}^k
    (-1)^{j + k' + l} 2^{j - 2 j'}
    \frac{N!}{(N - j - k - l + j' + k')!} \,
    \frac{l!}{(l - k')!} \,
    \frac{k!}{k'! \, (k - k')!} \,
    \frac{j!}{j'! \, (j - 2 j')!}
    \nonumber \\ & \qquad
    (\cosh^{N - j + 2 j'} \chi) (\sinh^{j - 2 j'} \chi)
    (\cos^{N - j - k - l + 2 j' + 2 k'} \zeta) (\sin^{k + l - 2 k'} \zeta)
    \, e^{- i \, (N - j - k - l) \, \theta + i \, (k - l) \, \varphi}
  \, ,
  \label{eq:f-phi:jkl}
\end{alignat}
where, for a positive integer $k$,
\[
  L_{\tensor{X}}^k
  := \underbrace{
      L_{\tensor{X}} \circ \cdots \circ L_{\tensor{X}}
    }_{k \textrm{ folds}}
\]
stands for the action of the Lie derivative $L_{\tensor{X}}$ for $k$ times,
and, $L_{\tensor{X}}^0$ (for $k = 0$) stands for the identity.
In the summation (\ref{eq:f-phi:jkl}), $\lfloor\frac{j}{2}\rfloor$ is the floor
of $\frac{j}{2}$, defined to be the largest integer less than or equal to
$\frac{j}{2}$.  Obviously,
\begin{equation}
  \phi_{\lambda}^{(000)} = \phi_\lambda
  \, .
\end{equation}
Here and after, we use the convention that $n! = \infty$ for negative integer
$n$.

According to the reprentation theory of Lie algebras \cite{Humphreys},
the complex vector space $V(\lambda)$ spanned by the functions
$
  L_{\tensor{f}_{\alpha_1 + \alpha_2}}^j
  L_{\tensor{f}_{\alpha_1 + 2 \alpha_2}}^k
  L_{\tensor{f}_{\alpha_1}}^l L_{\tensor{f}_{\alpha_2}}^n \phi_\lambda$
is an $\mathfrak{L}$-module, with the maximal vector $\phi_\lambda$ belonging to
the highest weight $\lambda = N \lambda_1$.  In the representation theory of
Lie algebras, such an $\mathfrak{L}$-module is called a Verma module.

It is easy to check, by virtue of eq.~(\ref{eq:f2:U}), that
\begin{equation}
  L_{\tensor{f}_{\alpha_2}} \phi_\lambda
  = \tensor{f}_{\alpha_2} \phi_\lambda
  = 0
  \, .
\end{equation}
In fact, this can be obviously seen from the shape of the weight diagram
(see, \S\ref{sect:weight-diagram}).
Then, it follows that the Verma module $V(\lambda)$ is spanned by the functions
$\phi_\lambda^{(jkl)}$ with nonnegative integers $j$, $k$ and $l$.

Because of $(-1)! = \infty$ and so on, we see from eq.~(\ref{eq:f-phi:jkl})
that, in order $ \phi_{\lambda}^{(jkl)} $ to be nonvanishing, all the following
conditions must be satisfied:
\begin{alignat}{1}
  & 0 \leqslant k' \leqslant k
  \, ,
  \label{phi:cond-1}
  \\
  & 0 \leqslant k' \leqslant l
  \, ,
  \label{phi:cond-2}
  \\
  & 0 \leqslant j' \leqslant \Big\lfloor \frac{j}{2} \Big\rfloor
  \, ,
  \label{phi:cond-3}
  \\
  & N - j - k - l + j' + k' \geqslant 0
  \, .
  \label{phi:cond-4}
\end{alignat}
Condition~(\ref{phi:cond-4}) is equivalent to
$ l \leqslant N - (j - j') - (k - k') $.
Combined with conditions~(\ref{phi:cond-1}) and (\ref{phi:cond-2}), this yields
\begin{equation}
  0 \leqslant k \leqslant N
  \, .
  \label{phi:cond:k}
\end{equation}
Similarly, we can obtain
\begin{equation}
  0 \leqslant l \leqslant N
  \, .
  \label{phi:cond:l}
\end{equation}
Conditions~(\ref{phi:cond-1}) and (\ref{phi:cond-2}) can be merged into
$ 0 \leqslant k' \leqslant \min(k, l) $.
It follows that
\[
  - k - l + k' \leqslant - \max(k, l)
  \, .
\]
So, condition~(\ref{phi:cond-4}) results in
\[
  j - j' \leqslant N - k - l + k' \leqslant N - \max(k, l)
  \, .
\]
On the other hand, from condition~(\ref{phi:cond-3}) we have
$ j - \lfloor \frac{j}{2} \rfloor \leqslant j - j' $.  Hence
\[
  j - \Big\lfloor \frac{j}{2} \Big\rfloor \leqslant N - \max(k, l)
  \, .
\]
No matter $ j \geqslant 0 $ is odd or even, there is always
$ \frac{j}{2} \leqslant j - \lfloor \frac{j}{2} \rfloor $.  Therefore,
\begin{equation}
  0 \leqslant j \leqslant 2 N - 2 \max(k, l)
  \, .
  \label{phi:cond:j}
\end{equation}
The inequalities~(\ref{phi:cond:l}), (\ref{phi:cond:k}) and (\ref{phi:cond:j})
are necessary and sufficient condition for $ \phi_{\lambda}^{(jkl)} $
to be nonzero.
These inequalities can be easily obtained with the aid of our knowledge of
the weight diagram (see, \S\ref{sect:weight-diagram}).

A corollary of the inequality~(\ref{phi:cond:j}) is very important:
since $ j + k + l \leqslant j + 2 \max(k, l) \leqslant 2 N$, we have
\begin{equation}
  \phi_\lambda^{(jkl)} = 0
  \, , \qquad (\textrm{whenever } j + k + l > 2 N)
  \, .
\end{equation}
This means that the Verma module $V(\lambda)$ is finite dimensional, coinciding
with conclusions in the representation theory of Lie algebras.

In appendix~\ref{sect:irreducibility} we shall prove that, for each integer
$N \geqslant 0$, the Verma module $V(\lambda)$ is an irreducible
$\mathfrak{L}$-module.

\subsection{Smoothness of $ \phi^{(jkl)}_{\lambda} $}

Strictly speaking, the function $\phi_\lambda^{(jkl)}$ in
eq.~(\ref{eq:f-phi:jkl}) is not defined globally on $dS^4$:  its domain is
just $U$, the union of coordinate neighborhoods of
$(\chi, \zeta, \theta, \varphi)$
defined by eqs.~(\ref{eq:coord-sys:U:0}) to (\ref{eq:coord-sys:U:4}).
In fact, we can use eqs.~(\ref{eq:coord-sys:U:0}) to (\ref{eq:coord-sys:U:4})
to obtain the line element on $U$,
$\od s^2 = g_{\mu\nu} \, \od x^\mu \, \od x^\nu$,
where $x^\mu$ for $\mu = 0$, \ldots, $3$ are $\chi$, $\zeta$, $\theta$ and
$\varphi$, respectively.  From eqs.~(\ref{eq:coord-sys:U:0}) to
(\ref{eq:coord-sys:U:4}) we can also obtain the invariant volume 4-form
\begin{equation}
  \vect{\omega} \at{U} = l^4 \, \cosh^3 \chi \sin \zeta \cos \zeta \;
    \od \chi \wedge \od \zeta \wedge \od \theta \wedge \od \varphi
    \, ,
  \label{eq:volume}
\end{equation}
indicating that
\begin{equation}
  \sqrt{- g} = l^4 \, \cosh^3 \chi \, | \sin \zeta \cos \zeta |
  \, ,
\end{equation}
where $g = \det (g_{\mu\nu})$.
It follows eq.~(\ref{eq:volume}) that, the functions $\chi$, $\zeta$, $\theta$
and $\varphi$ are not coordinates where $\sin \zeta = 0$ or $\cos \zeta = 0$.
Referring to eqs.~(\ref{eq:coord-sys:U:0}) to (\ref{eq:coord-sys:U:4}), we see
that $\cos\zeta = 0$ corresponds to $\xi^1 = \xi^2 = 0$, and that
$\sin\zeta = 0$ corresponds to $\xi^3 = \xi^4 = 0$.  These are two 2-surfaces
in $dS^4$.  Therefore the coordinate system $(\chi, \zeta, \theta, \varphi)$
does not cover them.

Now that the region $U$ in the above is $dS^4$ with the above two 2-surfaces
removed, it has four connected components, each of which is homeomorphic to
$\mathbb{R}^2 \times T^2$, with $T^2 = S^1 \times S^1$ the 2-torus.
Considering the periodicity of $\theta$ and $\varphi$,
none of the connected components is a genuine coordinate system, in fact:
each connected component of $U$ must be covered by at least four coordinate
neighborhood, on which $(\chi, \zeta, \theta, \varphi)$ is defined.
For details of these coordinate neighborhoods, we refer to the end of
\S\ref{sect:coord}.

On summary, the functions defined in eq.~(\ref{eq:f-phi:jkl}) are not globally
defined, so far.  In the following we shall show that each of them can be
``glued'' into a globally defined smooth function on $dS^4$.

Our strategy is as this:  because $\tensor{f}_\beta$ for each positive root
$\beta$ is a smooth vector field on $dS^4$, we need only to prove $\phi_\lambda$
is (or, can be ``glued'' into) a globally defined smooth function on $dS^4$.

We can define a function
\begin{equation}
  \widetilde{\phi}_\lambda
  = \bigg( \frac{\xi^1 - i \, \xi^2}{\sqrt{- \eta_{AB} \, \xi^A \xi^B}} \bigg)^N
  \label{eq:tilde-phi-lambda}
\end{equation}
on $\mathbb{R}^{1,4} - \{ \tensor{0} \}$, where $\tensor{0}$ is the zero vector
(the origin) of $\mathbb{R}^{1,4}$.
Obviously, $\widetilde{\phi}_\lambda$ is a smooth complex-valued function, and 
its restriction to $U$ is just $\phi_\lambda$
in eq.~(\ref{eq:phi-lambda})\footnote{
Equivalently, $ \phi_{\lambda} $ is the pullback
$i^* \widetilde{\phi}_\lambda$ of $\widetilde{\phi}_\lambda$,
where $i \colon dS^4 \hookrightarrow \mathbb{R}^{1,4}$ is the inclusion.
}.
This indicates that $\phi_\lambda$ is, in fact, a smooth function on $dS^4$.

So far we have proved that $\phi_\lambda$ is a smooth function on $dS^4$.
Hence so are $\phi_\lambda^{(jkl)}$.  As a consequence, $V(\lambda)$,
the vector space spanned by $\phi_\lambda^{(jkl)}$ with all integers $j$, $k$
and $l \geqslant 0$, is an $\mathfrak{L}$-submodule of $C^\infty(dS^4)$.

For the sake of later use, we give the functions
\begin{alignat}{1}
  \widetilde{\phi}_\lambda^{(jkl)}
  &= \sum_{j' = 0}^{\lfloor\frac{j}{2}\rfloor} \sum_{k' = 0}^k
    (-1)^{j + k' + l} e^{j - 2 j'} \,
    \frac{N!}{(N - j - k - l + j' + k')!} \,
    \frac{l!}{(l - k')!} \,
    \frac{k!}{k'! \, (k - k')!} \,
    \frac{j!}{j'! \, (j - 2 j')!}
    \nonumber \\ & \qquad
    \frac{
      (\xi^0)^{j - 2 j'}
      (\xi^1 - i \, \xi^2)^{N - j - k - l + j' + k'}
      (\xi^1 + i \, \xi^2)^{j' + k'}
      (\xi^3 - i \, \xi^4)^{l - k'} (\xi^3 + i \, \xi^4)^{k - k'}
    }{(- \eta_{AB} \, \xi^A \xi^B)^{\frac{N}{2}}}
\end{alignat}
for integers $j$, $k$ and $l \geqslant 0$.
Each of such functions is a smooth function on
$\mathbb{R}^{1,4} - \{ \tensor{0} \}$.
The restriction of $\widetilde{\phi}_\lambda^{(jkl)}$ to $dS^4$ is just
$\phi_\lambda^{(jkl)}$ in eq.~(\ref{eq:f-phi:jkl}).  Equivalently,
\begin{equation}
  i^* \widetilde{\phi}_\lambda^{(jkl)} = \phi_\lambda^{(jkl)}
  \, .
\end{equation}
This clearly shows that $\phi_\lambda^{(jkl)}$ is a smooth function on $dS^4$.

\subsection{Weight Spaces and Weight Diagram of $V(\lambda)$}
\label{sect:weight-diagram}

A dual vector $\mu \in \mathfrak{h}'^*$ is a linear function on $\mathfrak{h}'$,
mapping each $h \in \mathfrak{h}'$ to a number $\mu(h) \in \mathbb{C}$.
Any $\mu \in \mathfrak{h}'^*$ can be associated with a linear subspace
$V(\lambda)_\mu$ of $V(\lambda)$, consisting of all $\phi \in V(\lambda)$
satisfying
\begin{equation}
  L_{\tensor{h}} \phi = \tensor{h} \phi
  = \mu(h) \, \phi
  \, , \qquad
  \forall h \in \mathfrak{h}'
  \, .
\end{equation}
Ordinarily $V(\lambda)_\mu$ consists of only $0 \in V(\lambda)$, the zero
function on $dS^4$.
But, for certain $\mu \in \mathfrak{h}'^*$, the corresponding $V(\lambda)_\mu$
can be nontrivial.  In this case, every nonzero $\phi \in V(\lambda)_\mu$ 
is a common eigenvector of both $\tensor{h}_{\alpha_1}$ and
$\tensor{h}_{\alpha_2}$, hence all $\tensor{h}$ (corresponding to
$h \in \mathfrak{h}'$).  Such a dual vector $\mu \in \mathfrak{h}'^*$ is called
a weight of the $\mathfrak{L}$-module $V(\lambda)$, and $V(\lambda)_\mu$ is
called the weight space with the weight $\mu$.
The set $\Pi(\lambda)$, consisting of all weights of $V(\lambda)$, is called
the weight diagram of $V(\lambda)$.

For example, the smooth function $\phi_\lambda^{(jkl)}$ satisfies
\begin{equation}
  L_{\tensor{h}} \phi_\lambda^{(jkl)}
  = \mu(h) \, \phi_\lambda^{(jkl)}
  \, , \qquad
  \forall h \in \mathfrak{h}'
  \, .
\end{equation}
where
\begin{alignat}{1}
  \mu &= \lambda
  - j \, (\alpha_1 + \alpha_2)
  - k \, (\alpha_1 + 2 \, \alpha_2)
  - l \, \alpha_1
  = n_1 \, \lambda_1 + n_2 \, \lambda_2
  \label{eq:weight:mu}
\end{alignat}
with
\begin{equation}
  n_1 = N - j - 2 l
  \, , \qquad
  n_2 = -2 \, (k - l)
  \, .
  \label{eq:jkl-n1n2}
\end{equation}
On account of the construction of $V(\lambda)$, the following observation is
obvious:  For a fixed pair of integers $n_1$ and $n_2$ so that
$\mu = n_1 \lambda_1 + n_2 \lambda_2$ is a weight, $V(\lambda)_\mu$ is spanned
by all nonzero $\phi_\lambda^{(jkl)}$, in which nonnegative integers $j$, $k$
and $l$ satisfy eqs.~(\ref{eq:jkl-n1n2}).

\begin{figure}[hbt]
  \centering
  \begin{tabular}{ccc}
    \includegraphics{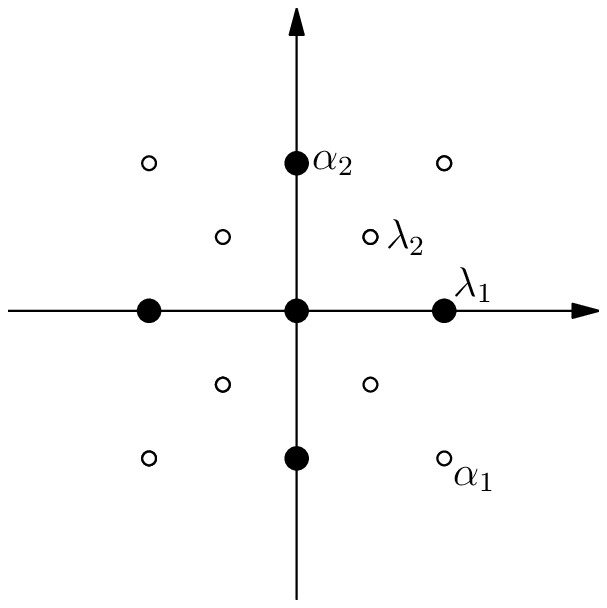}
    & \qquad &
    \includegraphics{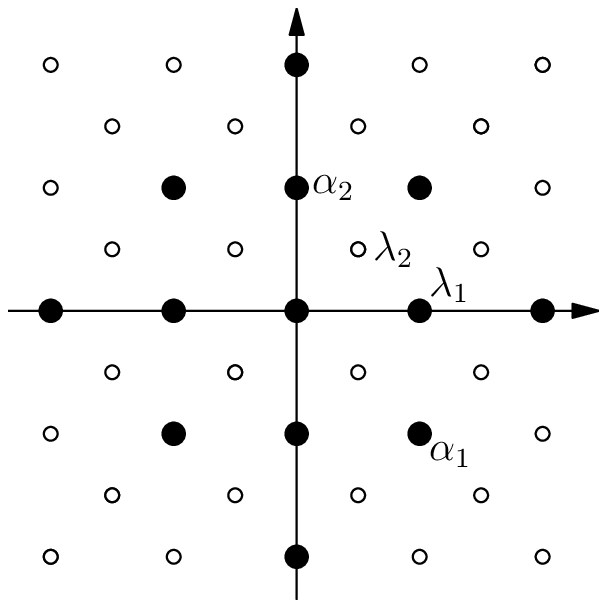}
    \\
    (a)~$\lambda = \lambda_{1}$
    &&
    (b)~$\lambda = 2 \lambda_{1}$
  \end{tabular}
  \caption{Weight diagrams with the highest weight $\lambda = \lambda_{1}$
    and $\lambda = 2 \lambda_{1}$.
    (Circles are not weights.  Only the dots in black are weights.)}
  \label{fig:weight-diagram-1}
\end{figure}

Using the representation theory of semisimple Lie algebras \cite{Humphreys},
the weight diagram $\Pi(\lambda)$ can be obtained, starting from the highest
weight $\lambda$.  In Figure~\ref{fig:weight-diagram-1}, we show
$\Pi(\lambda_1)$ and $\Pi(2 \lambda_2)$ as two examples.
One should pay attention that circles in the figures are not weights.
Then, the $\mathfrak{L}$-module $V(\lambda)$ can be decomposed into
the direct sum of weight spaces:
\begin{equation}
  V(\lambda) = \bigoplus_{\mu \in \Pi(\lambda)} V(\lambda)_\mu
  \, .
  \label{eq:decomposition:weight-space}
\end{equation}

In fact, from the knowledge of weight diagrams \cite{Humphreys}, we can label
the weights of $V(\lambda)$, where $\lambda = N \, \lambda_1$, one by one
as follows:  $\mu \in \Pi(\lambda)$ if and only if
\begin{equation}
  \mu = \lambda - l \, \alpha_1 - j \, (\alpha_1 + \alpha_2)
  \, , \qquad
  (0 \leqslant l \leqslant N
    \, , \quad
    0 \leqslant j \leqslant 2 N - 2 l)
  \label{eq:weight:upper}
\end{equation}
or
\begin{equation}
  \mu = \lambda - k \, (\alpha_1 + 2 \, \alpha_2) - j \, (\alpha_1 + \alpha_2)
  \, , \qquad
  (0 < k \leqslant N
    \, , \quad
    0 \leqslant j \leqslant 2 N - 2 k)
  \, .
  \label{eq:weight:down}
\end{equation}
In order for $\phi_\lambda^{(jkl)}$ to be nonzero, both $\phi_\lambda^{(00l)}$
and $\phi_\lambda^{(0kl)}$ must be nonzero.
Consequently, the necessary condition for $\phi_\lambda^{(jkl)}$ to be nonzero
is
\begin{equation}
  0 \leqslant l \leqslant N
  \, , \quad
  0 \leqslant k \leqslant N
  \, , \quad
  0 \leqslant j \leqslant 2 \min(N - k, N - l)
  \, .
  \label{condt:jkl}
\end{equation}
These conditions can also be obtained by analyzing the non-vanishing condition
for $\phi_\lambda^{(jkl)}$ in eq.~(\ref{eq:f-phi:jkl}).

Notice that $\lambda_1 = \alpha_1 + \alpha_2$ and $\alpha_2$ are orthogonal,
having the same length.  So it is often convenient to write $\mu$
in eq.~(\ref{eq:weight:mu}) as
\begin{equation}
  \mu 
  = \Big( n_1 + \frac{n_2}{2} \Big) \, \lambda_1 + \frac{n_2}{2} \, \alpha_{2}
  \, .
  \label{eq:weight:orthonormal-basis}
\end{equation}
Note that $n_2$ is an even integer.  (See, eqs.~(\ref{eq:jkl-n1n2}).)
Then the necessary and sufficient condition for $\mu$ to be a weight is
\begin{alignat*}{1}
  - N \leqslant \Big( n_1 + \frac{n_2}{2} \Big) + \frac{n_2}{2} \leqslant N
  \, , \qquad
  - N \leqslant \Big( n_1 + \frac{n_2}{2} \Big) - \frac{n_2}{2} \leqslant N
  \, .
\end{alignat*}
That is, the integer $n_1$ and the even integer $n_2$ must satisfy the following
inequalities:
\begin{equation}
  - N \leqslant n_1 \leqslant N
  \, , \qquad
  - N \leqslant n_1 + n_2 \leqslant N
  \, .
  \label{weight-diagram:n1n2}
\end{equation}
This can be directly seen from the figure of $\Pi(\lambda)$, and can be derived
from the label (\ref{eq:weight:upper}) and (\ref{eq:weight:down}).

\subsection{Multiplicity of Weights}
\label{sect:multiplicity}

One of the consequence of eq.~(\ref{eq:decomposition:weight-space}) is
\begin{equation}
  \dim V(\lambda) = \sum_{\mu \in \Pi(\lambda)} \dim V(\lambda)_\mu
  \, .
  \label{eq:dim:V(lambda)}
\end{equation}
Traditionally $\dim V(\lambda)_\mu$ is called the multiplicity of $\mu$
with respect to the highest weight $\lambda$, and denoted by $m_\lambda(\mu)$.
In Appendix~\ref{sect:proof:multiplicity} we shall prove that, for a weight
$\mu = n_1 \, \lambda_1 + n_2 \, \lambda_2$,
\begin{equation}
  m_\lambda(n_1 \, \lambda_1 + n_2 \, \lambda_2)
  = \Big\lfloor
      \frac{N - |n_1 + \frac{n_2}{2}| - |\frac{n_2}{2}|}{2}
    \Big\rfloor
  + 1
  \, ,
  \label{eq:multiplicity}
\end{equation}
where $\lfloor \cdot \rfloor$ denotes the floor of a number.
Equivalently, for a weight $l_1 \, \lambda_1 + l_2 \, \alpha_2$, we have
\begin{equation}
  m_\lambda (l_1 \, \lambda_1 + l_2 \, \alpha_2)
  = \Big\lfloor \frac{N - |l_1| - |l_2|}{2} \Big\rfloor + 1
  \, .
\end{equation}
See, eq.~(\ref{eq:weight:orthonormal-basis}).

\subsection{Linear Dependence}

Given a weight $\mu = n_1 \, \lambda_1 + n_2 \, \lambda_2$, the weight space
$V(\lambda)_\mu$ is spanned by the functions $\phi_\lambda^{(jkl)}$ with
nonnegative integers $j$, $k$ and $l$ satisfying eq.~(\ref{eq:weight:mu}), or
equivalently, eqs.~(\ref{eq:jkl-n1n2}).  Then, arises a question:
are all the nonzero functions $ \phi^{(jkl)}_{\lambda} $ linearly independent?

The answer is negative.  For example, when $ N = 1 $, nonzero functions
$ \phi^{(jkl)}_{\lambda} $ are listed as follows.
\begin{alignat}{1}
  \phi^{(000)}_{\lambda_{1}} &= \cosh \chi \cos \zeta \, e^{-i \theta}
  \, , \\
  \phi^{(001)}_{\lambda_{1}} &= - \cosh \chi \sin \zeta \, e^{- i \varphi}
  \, , \\
  \phi^{(010)}_{\lambda_{1}} &= \cosh \chi \sin \zeta \, e^{i \varphi}
  \, , \\
  \phi^{(011)}_{\lambda_{1}} &= \cosh \chi \cos \zeta \, e^{i \theta}
  \, , \\
  \phi^{(100)}_{\lambda_{1}} &= - 2 \sinh \chi
  \, , \\
  \phi^{(200)}_{\lambda_{1}} &= 2 \cosh \chi \cos \zeta \, e^{i \theta}
  \, .
\end{alignat}
Obviously, $ \phi^{(011)}_{\lambda_{1}} $ and $ \phi^{(200)}_{\lambda_{1}} $
are linearly dependent.

\subsection{Nonexistence of Infinite Dimensional $\mathfrak{L}$-Submodules in
  $C^{\infty}(dS^{4})$}

So far the Verma modules contained in $C^{\infty}(dS^{4})$ are all irreducible
and finite dimensional.  There is a question, then, whether there exist any
infinite dimensional irreducible $\mathfrak{L}$-submodules
of $C^{\infty}(dS^{4})$, where
$\mathfrak{L} = \mathfrak{so}(1, 4) \otimes_{\mathbb{R}} \mathbb{C}$.
The answer is negative.

In \S\ref{sect:Verma-modules}, the Verma modules are constructed according to
the representation theory\cite{Humphreys} of semisimple Lie algebras:
We first start from a dominant weight $\lambda$ as the highest weight.
Then, in the process of determining the maximal vector $\phi_{\lambda}$,
the highest weight is also determined to be $N \lambda_{1}$, with $N$
a nonnegative integer.

In \S\ref{sect:Verma-modules}, the reason for $N$ to be a nonnegative integer
comes from the representation theory of semisimple Lie algebras:
an irreducible highest weight module is finite dimensional if and only if
its highest weight is dominant and integral~\cite{Humphreys}.

In fact, in \S\ref{sect:Verma-modules} we need not refer to the representation
theory, just remaining $N$ in eq.~(\ref{eq:phi-lambda}) to be an unknown
parameter.  Now that the maximal vector $\phi_{\lambda}$ has been determined
in the form of eq.~(\ref{eq:phi-lambda}), the parameter $N$ must be an integer
because of the periodicity of $\theta$.  See, eqs.~(\ref{eq:coord-sys:U:1}) and
(\ref{eq:coord-sys:U:2}).  If $N$ is negative, however, the function
$\widetilde{\phi}_{\lambda}$ in eq.~(\ref{eq:tilde-phi-lambda}) is not a smooth
function on $\mathbb{R}^{1, 4} - \{ \tensor{0} \}$.  Since the pullback (or,
naively, restriction) of $\widetilde{\phi}_{\lambda}$ to $U \subset dS^{4}$
is just $\phi_{\lambda}$, the latter cannot be extended to be a smooth function
on $dS^{4}$ provided $N < 0$.  Hence, without referring to the representation
theory of semisimple Lie algebras, we can still determine that $N$ is
a nonnegative integer.  Hence, a Verma module $V(N \lambda_{1})$ is always
finite dimensional and irreducible.

In one word, an irreducible $\mathfrak{L}$-submodule of $C^{\infty}(dS^{4})$
is always finite dimensional, being a Verma module $V(N \lambda_{1})$
with $N$ a nonnegative integer.

\section{Smooth Solutions and Masses of the Klein-Gordon Scalars on $dS^4$}
\label{sect:solution-mass}

\subsection{Imaginary and Discrete Masses of the Klein-Gordon Equation
  on $ dS^{4} $}

It is well known that
\begin{equation}
  C = \frac{1}{2} \, \eta^{AC} \eta^{BD} \, X_{AB} X_{CD}
\end{equation}
is a universal Casimir element of
$\mathfrak{L} = \mathfrak{so}(1, 4) \otimes_{\mathbb{R}} \mathbb{C}$.
When acting on the irreducible $\mathfrak{L}$-module $V(\lambda)$,
where $\lambda = N \lambda_1$, $X_{AB}$ is replaced by the Lie derivative
$L_{\tensor{X}_{AB}}$, or directly, the vector field $\tensor{X}_{AB}$.
So is $h_{\alpha_1}$, $e_{\alpha_1}$, $f_{\alpha_1}$, and so on,
in the following.  Using the expressions (\ref{eq:Chevalley-1:II}) to
(\ref{eq:Chevalley-5:II}), we can verify that
\begin{alignat}{1}
  C &= - \frac{1}{2} \, h_{\alpha_1} (2 h_{\alpha_1} + h_{\alpha_2})
  - \frac{1}{2} \, h_{\alpha_2} (h_{\alpha_1} + h_{\alpha_2})
  - (e_{\alpha_1} f_{\alpha_1} + f_{\alpha_1} e_{\alpha_1})
  - \frac{1}{2} \, (e_{\alpha_2} f_{\alpha_2} + f_{\alpha_2} e_{\alpha_2})
  \nonumber \\ &
  - \frac{1}{2} \, (
      e_{\alpha_1 + \alpha_2} f_{\alpha_1 + \alpha_2}
      + f_{\alpha_1 + \alpha_2} e_{\alpha_1 + \alpha_2}
    )
  - (
      e_{\alpha_1 + 2 \alpha_2} f_{\alpha_1 + 2 \alpha_2}
      + f_{\alpha_1 + 2 \alpha_2} e_{\alpha_1 + 2 \alpha_2}
    )
  \, .
\end{alignat}
First using the commutators in Appendix~\ref{sect:commutators}, then using
eqs.~(\ref{eq:h:alpha+}), we can reduce the above expression to
\begin{alignat}{1}
  C &= - h_{\alpha_{1}} h_{\alpha_{1}}
  - h_{\alpha_{1}} h_{\alpha_{2}}
  - \frac{1}{2} \, h_{\alpha_{2}} h_{\alpha_{2}}
  - 3 \, h_{\alpha_{1}}
  - 2 \, h_{\alpha_{2}}
  \nonumber \\ & \phantom{={}}
  - 2 \, f_{\alpha_{1}} e_{\alpha_{1}}
  - f_{\alpha_{2}} e_{\alpha_{2}}
  - f_{\alpha_{1} + \alpha_{2}} e_{\alpha_{1} + \alpha_{2}}
  - 2 \, f_{\alpha_{1} + 2 \alpha_{2}} e_{\alpha_{1} + 2 \alpha_{2}}
  \, .
\end{alignat}
When $ C $ acts on $ \phi_{\lambda} $, there is simply
\begin{alignat*}{1}
  C. \phi_{\lambda}
  &= - h_{\alpha_{1}} h_{\alpha_{1}} . \phi_{\lambda}
  - h_{\alpha_{1}} h_{\alpha_{2}} . \phi_{\lambda}
  - \frac{1}{2} \, h_{\alpha_{2}} h_{\alpha_{2}} . \phi_{\lambda}
  - 3 \, h_{\alpha_{1}} . \phi_{\lambda}
  - 2 \, h_{\alpha_{2}} . \phi_{\lambda}
  \nonumber \\
  &= - [\lambda(h_{\alpha_{1}})]^{2} \phi_{\lambda}
  - \lambda(h_{\alpha_{2}}) \, \lambda(h_{\alpha_{1}}) \, \phi_{\lambda}
  - \frac{1}{2} \, [\lambda(h_{\alpha_{2}})]^{2} \phi_{\lambda}
  - 3 \, \lambda(h_{\alpha_{1}}) \, \phi_{\lambda}
  - 2 \, \lambda(h_{\alpha_{2}}) \, \phi_{\lambda}
  \, .
\end{alignat*}
Note that $\lambda(h_{\alpha_1}) = N \, \lambda_1 (h_{\alpha_1}) = N$ and
$\lambda(h_{\alpha_2}) = N \, \lambda_1 (h_{\alpha_2}) = 0$.
Hence
\begin{alignat*}{1}
  C. \phi_\lambda
  &= - N^2 \, \phi_\lambda
  - 3 N \, \phi_\lambda
  = - N \, (N + 3) \, \phi_\lambda
  \, .
\end{alignat*}
Since $ V(\lambda) $ is an irreducible $ \mathfrak{L} $-module,
according to Schur's lemma, every $\phi \in V(\lambda)$ satisfies
\begin{equation}
  C. \phi = - N \, (N + 3) \, \phi
  \, .
\end{equation}
Comparing it with eq.~(\ref{eq:KG:Casimir}), we have the mass $m$
of the Klein-Gordon field $\phi$, as shown in the following:
\begin{equation}
  m^2 = - \frac{N (N + 3) \, \hbar^2}{c^2 l^2}
  \, .
  \label{eq:mass-squared}
\end{equation}

It is significant that the mass is not only discrete, but also an imaginary
quantity.  In the classical level, this doesn't matter, because $m$ only
makes sense in the quantum level.  The detailed consequence and discussion of
this fact in QFT will be presented in other papers.

\subsection{Irreducibility of the Solution Space of a Klein-Gordon Equation}

So far we have shown that the Klein-Gordon equation on $dS^4$ must be of
the form
\begin{equation}
  g^{ab} \, \nabla_{a} \nabla_{b} \phi - \frac{N \, (N + 3)}{l^2} \, \phi
  = 0
  \label{eq:KG:N}
\end{equation}
with certain a nonnegative integer $N$.  We have shown that each smooth
function $\phi \in V(N \lambda_1)$ is a solution of the above equation.
That is, the irreducible $\mathfrak{L}$-module $V(N \lambda_1)$ is a linear
subspace of the solution space $ \mathscr{S}_{\textrm{KG}}(dS^{4}) $ of
eq.~(\ref{eq:KG:N}).  Since $ \mathscr{S}_{\textrm{KG}}(dS^{4}) $ is also
an $\mathfrak{L}$-module, but not necessarily irreducible,
it must be the direct sum of $V(N' \lambda_1)$ with $N' \geqslant 0$ satisfying
\[
  N' \, (N' + 3) = N \, (N + 3)
  \, .
\]
It is easy to check that the only possibility is $N' = N$.  Consequently,
the solution space of eq.~(\ref{eq:KG:N}) is $V(N \lambda_1)$, the irreducible
$\mathfrak{L}$-module having $N \lambda_1$ as its highest weight.

So, a general smooth solution of the Klein-Gordon equation on $ dS^{4} $,
namely, eq.~(\ref{eq:KG:N}), is
\begin{equation}
  \phi = \sum_{l = 0}^{N} \sum_{k = 0}^{N} \sum_{j = 0}^{2 N - 2 \max(k, l)}
    a_{jkl} \, \phi^{(jkl)}_{N \lambda_{1}}
  \, ,
\end{equation}
where $ a_{jkl} $ are some complex constants.  Although the functions
$ \phi^{(jkl)}_{N \lambda_{1}} $ are possibly linearly dependent, the conclusion
remains true, only that the coefficients $ a_{jkl} $ for a given solution
are not uniquely determined.


\section{Conclusions and Discussion}
\label{sect:CD}

There are papers, such as \cite{YG} and \cite{KV}, discussing the solutions of
the Klein-Gordon equation on $dS^{4}$.  The discussion in \cite{YG} does not
evaluate the effects influenced by the global structures of the spacetime.
Note that the elegant method in \cite{ChangGuo} can be applied to $dS^{4}$,
too.  But the solutions obtained in \cite{ChangGuo} are massless scalars,
and the smoothness of these solutions were not discussed.
By imposing the condition of quadratically integrable on the whole $dS^{4}$,
it is shown in \cite{KV} that the mass $m$ of Klein-Gordon fields on $dS^{4}$
satisfies (in the natural units)
\[
  m^{2} = \lambda_{jn}
  = \frac{9}{4} - \Big(j + \frac{1}{2} - n\Big)^{2}
\]
with $j$, $n = 0$, 1, 2, \ldots such that $j + \frac{1}{2} - n > 0$.
If we set $N = j - n - 1$, there will be $N > - \frac{3}{2}$ and
$m^{2} = - N (N + 3)$.  In Theorem~{2} in \cite{KV} it is stated that
$m^{2}$ could be positive (then equal to 2), and that the solution space
for each $\lambda_{jn}$ is infinite dimensional.  Although the mass spectrum
is very similar to ours, but in some details, the conclusions are quite
different.  Since there is no detailed proof in \cite{KV}, this will be left
as an open question.

By using the Lie group and Lie algebra method, we have obtained all smooth
solutions of a Klein-Gordon equation in the de~Sitter background,
forming a finite dimensional irreducible $\mathfrak{L}$-module,
with $\mathfrak{L} = \mathfrak{so}(1, 4) \otimes_{\mathbb{R}} \mathbb{C}$.
An associated conclusion is that the mass of a Klein-Gordon equation on $dS^{4}$
cannot be arbitrary.  It's square must be non-positive and discrete, as shown
in eq.~(\ref{eq:mass-squared}).

In this paper we construct the irreducible $\mathfrak{L}$-modules with respect
to the Cartan subalgebra $\mathfrak{h}'$, spanned by $X_{12}$ and $X_{34}$.
Coordinate systems and weight spaces can be constructed with respect to
the Cartan subalgebra $\mathfrak{h}$, spanned by $X_{12}$ and $X_{04}$.
Detailed discussion will be presented in other papers.

So far, it is not so factory that the functions $ \phi^{(jkl)}_{N \lambda_{1}} $
(with $ j $, $ k $ and $ l $ satisfying the condition (\ref{condt:jkl}))
might be not linearly independent.  But the details of a solution is not our
main topic in this paper.  These are left for future papers.  As a consequence,
it is not quite suitable for now to discuss the quantization of Klein-Gordon
fields in the de~Sitter background.

But problem due to the mass must be discussed here.
When viewed as a relativistic quantum mechanical equation,
the Klein-Gordon equation in the Minkowski background is obtained
by applying the quantization rule
\[
  p_{\mu} \to \hat{p}_{\mu} = i \hbar \frac{\partial}{\partial x^{\mu}}
\]
to the relation
$ \eta_{\mu\nu} \, p^{\mu} p^{\nu} = m^{2} c^{2} $.
Then, from the Klein-Gordon equation in the Minkowski background to that
in a curved spacetime, we need only to replace the partial derivative
to the covariant derivative.
Unfortunately, in the case of de~Sitter background, we have seen that the mass
is no longer real (except when $ N = 0 $).

What if we exchange these two steps: first establishing the classical mechanics
in the de~Sitter background, then quantizing it?
H.-Y. Guo \textit{et al} have attempted the first step:
trying their best to establish a classical mechanics resembling the relativistic
mechanics.  For a free particle in $ dS^{4} $, there exists a conserved
5-angular momentum $ \mathcal{L}_{AB} $, satisfying the equality
\begin{equation}
  - \frac{1}{2 l^{2}} \, \eta^{AC} \eta^{BD} \,
    \mathcal{L}_{AB} \mathcal{L}_{CD}
  = \frac{E^{2}}{c^{2}} - \vect{P}^{2} - \frac{1}{2 l^{2}} \, \vect{L}^{2}
  = m_{\Lambda 0}^{2} c^{2}
  \, ,
  \label{eq:L-m}
\end{equation}
where $ E $, $ \vect{P} $ and $\vect{L} $ are the splitting of
the 5-angular momentum with respect to a Beltrami coordinate system,
and $ m_{\Lambda 0} $ is the proper mass of
the particle~\cite{BdS1, BdS2, BdS3, BdS4}.
When $ l \to \infty $, $ (E/c, \vect{P}) $ tends to the 4-momentum
in Einstein's special relativity, while $ \vect{L}/l \to 0 $.
Under a reasonable quantization rule
\begin{equation}
  \mathcal{L}_{AB} \rightarrow i \hbar \, \tensor{X}_{AB}
  \, ,
\end{equation}
the above equation will yield a ``quantum'' equation
\[
  \frac{\hbar^{2}}{2 l^{2}} \, \eta^{AC} \eta^{BD} \,
    L_{\tensor{X}_{AB}} L_{\tensor{X}_{CD}} \phi
  = m_{\Lambda 0}^{2} c^{2} \, \phi
  \, ,
\]
namely, the Klein-Gordon equation~(\ref{eq:KG}).
Unfortunately still, in the classical level, i.e., in eq.~(\ref{eq:L-m}),
the mass $ m_{\Lambda 0} $ is nonnegative, while in the resulted ``quantum''
equation, $ m_{\Lambda 0}^{2} \leqslant 0 $.

This really sounds bad, because the process of quantization and the process of
generalizing to curved spacetime seems not so compatible.
For long there are some physicists believing that general relativity and
quantum theory are not compatible.  Even if they were wrong eventually,
this problem is at least very serious and hard currently:
before we settled down to investigation of QFT in the de~Sitter background,
we must suitably solve the problem of $ m^{2} \leqslant 0 $.

Another belief is, when the cosmological radius $ l \to \infty $ in $ dS^{4} $,
physical laws and phenomena tend to those in the Minkowski spacetime.
The Klein-Gordon equation is again an exception:
On the one hand, the problem of $ m^{2} \leqslant 0 $ is still the obstacle.
On the other hand, the dimension of the solution space is also an obstacle,
with the one for the Minkowski space being infinite dimensional, while the one
for $dS^{4}$ being finite dimensional.

At last, we point out that the method in this paper can be applied to various
field equations in de~Sitter spacetime or anti-de~Sitter spacetime.  These will
be presented in other papers.  For other spacetimes with sufficient symmetries,
this method might be effective, too.

\section*{\centering Acknowledgment}

We are grateful to Prof. Yongge~Ma for helpful and critical discussions.
The first author wants to express his special thanks to Professors
Zhan~Xu, Chao-Guang~Huang, Yu~Tian, Xiaoning~Wu for the continuing
long term cooperation, and for stimulations during the cooperation.
He also had many in-depth discussions with Prof. Huai-Yu~Wang.
During this work, which lasts for
quite a long time, the first author had many helpful discussions with
Professors Rong-Gen~Cai, Rong-Jia~Yang, Yang~Zhang, Xuejun~Yang and
Dr. Hong-Tu~Wu, Chun-Liang~Liu, Shibei~Kong and Wei~Zhang.
And, finally, the first author owes much to the late Prof. Han-Ying~Guo.

This work is partly supported by the Fundamental Research Funds for the Central
Universities under the grant No.~105116.

\appendix

\section{The Commutators of
  $\mathfrak{so}(1, 4) \otimes_{\mathbb{R}} \mathbb{C}$}
\label{sect:commutators}

The formulae in this section are satisfied, no matter the Cartan subalgebra
is $ \mathfrak{h} $ in \S\ref{sect:h} or $ \mathfrak{h}' $ in \S\ref{sect:h'}.

First of all, for each $\alpha$ and $\beta \in \Phi^+$, there are the standard
commutators
\begin{equation}
  [h_\alpha, e_\beta] = \beta(h_{\alpha}) \, e_\beta
  \, , \qquad
  [h_\alpha, f_\beta] = - \beta(h_{\alpha}) \, f_\beta
  \, , \qquad
  [e_\beta, f_\beta] = h_\beta
  \, .
\end{equation}
Note that
\begin{equation}
  h_{\alpha_1 + \alpha_2} = 2 \, h_{\alpha_1} + h_{\alpha_2}
  \, , \qquad
  h_{\alpha_1 + 2 \alpha_2} = h_{\alpha_1} + h_{\alpha_2}
  \, .
  \label{eq:h:alpha+}
\end{equation}
Thus, $\beta(h_{\alpha})$ for all $\alpha$ and $\beta \in \Phi^+$ can be
obtained by virtue of the linear property of $\beta$ and the Cartan integers
\begin{equation}
  \alpha_i (h_{\alpha_j}) = \langle \alpha_i, \alpha_j \rangle
\end{equation}
for $i$, $j = 1$ and $2$.

Next, the following commutators are satisfied by the Chevalley bases
in both \S\ref{sect:h} and \S\ref{sect:h'}:
\begin{alignat}{3}
  [e_{\alpha_1}, e_{\alpha_2}] &= e_{\alpha_1 + \alpha_2}
  \, , &
  [e_{\alpha_1}, e_{\alpha_1 + \alpha_2}] &= 0
  \, , &
  [e_{\alpha_1}, e_{\alpha_1 + 2 \alpha_2}] &= 0
  \, , \\
  [e_{\alpha_2}, e_{\alpha_1 + \alpha_2}] &= 2 \, e_{\alpha_1 + 2 \alpha_2}
  \, , &
  [e_{\alpha_2}, e_{\alpha_1 + 2 \alpha_2}] &= 0
  \, , &
  [e_{\alpha_1 + \alpha_2}, e_{\alpha_1 + 2 \alpha_2}] &= 0
  \, , \\
  [f_{\alpha_1}, f_{\alpha_2}] &= - f_{\alpha_1 + \alpha_2}
  \, , &
  [f_{\alpha_1}, f_{\alpha_1 + \alpha_2}] &= 0
  \, , &
  [f_{\alpha_1}, f_{\alpha_1 + 2 \alpha_2}] &= 0
  \, , \\
  [f_{\alpha_2}, f_{\alpha_1 + \alpha_2}] &= - 2 \, f_{\alpha_1 + 2 \alpha_2}
  \, , \qquad &
  [f_{\alpha_2}, f_{\alpha_1 + 2 \alpha_2}] &= 0
  \, , \qquad &
  [f_{\alpha_1 + \alpha_2}, f_{\alpha_1 + 2 \alpha_2}] &= 0
  \, , \\
  [e_{\alpha_1}, f_{\alpha_2}] &= 0
  \, , &
  [e_{\alpha_1}, f_{\alpha_1 + \alpha_2}] &= - f_{\alpha_2}
  \, , &
  [e_{\alpha_1}, f_{\alpha_1 + 2 \alpha_2}] &= 0
  \, , \\
  [e_{\alpha_2}, f_{\alpha_1}] &= 0
  \, , &
  [e_{\alpha_2}, f_{\alpha_1 + \alpha_2}] &= 2 f_{\alpha_1}
  \, , &
  [e_{\alpha_2}, f_{\alpha_1 + 2 \alpha_2}] &= - f_{\alpha_1 + \alpha_2}
  \, , \\
  [e_{\alpha_1 + \alpha_2}, f_{\alpha_1}] &= - e_{\alpha_2}
  \, , &
  [e_{\alpha_1 + \alpha_2}, f_{\alpha_2}] &= 2 \, e_{\alpha_1}
  \, , &
  [e_{\alpha_1 + \alpha_2}, f_{\alpha_1 + 2 \alpha_2}] &= f_{\alpha_2}
  \, , \\
  [e_{\alpha_1 + 2 \alpha_2}, f_{\alpha_1}] &= 0
  \, , &
  [e_{\alpha_1 + 2 \alpha_2}, f_{\alpha_2}] &= - e_{\alpha_1 + \alpha_2}
  \, , \quad &
  [e_{\alpha_1 + 2 \alpha_2}, f_{\alpha_1 + \alpha_2}] &= e_{\alpha_2}
  \, .
\end{alignat}

\section{Irreducibility of the Verma Module $V(\lambda)$}
\label{sect:irreducibility}

According to the representation theory of Lie algebras, the function
in eq.~(\ref{eq:f-phi:jkl}) belongs to the weight
\begin{alignat}{1}
  \mu &= \lambda
  - j \, (\alpha_1 + \alpha_2)
  - k \, (\alpha_1 + 2 \alpha_2)
  - l \, \alpha_1
  = N \, \lambda_1
  - (j + k + l) \, \alpha_1
  - (j + 2 k) \, \alpha_2
  \nonumber \\
  &= n_1 \, \lambda_1 + n_2 \, \lambda_2
  \, ,
\end{alignat}
where
\begin{equation}
  n_1 = N - j - 2 l
  \, , \qquad
  n_2 = - 2 \, (k - l)
  \, .
\end{equation}
That is, it satisfies the conditions
\begin{alignat}{1}
  h_{\alpha_1}.\phi_\lambda^{(jkl)}
  &= L_{\tensor{h}_{\alpha_1}} \phi_\lambda^{(jkl)}
  = \mu(h_{\alpha_1}) \, \phi_\lambda^{(jkl)}
  = n_1 \, \phi_\lambda^{(jkl)}
  \, , \\
  h_{\alpha_2}.\phi_\lambda^{(jkl)}
  &= L_{\tensor{h}_{\alpha_2}} \phi_\lambda^{(jkl)}
  = \mu(h_{\alpha_2}) \, \phi_\lambda^{(jkl)}
  = n_2 \, \phi_\lambda^{(jkl)}
  \, ,
\end{alignat}
which can be obviously seen from eqs.~(\ref{eq:f-phi:jkl}), (\ref{eq:h1:U}) and
(\ref{eq:h2:U}).

In order that $\phi_\lambda^{(jkl)} \neq 0$ in $V(\lambda)$ (meaning that
this function is nonzero somewhere on $dS^4$), there must be
$j + k + l \leqslant N$, namely,
\begin{equation}
  n_1 + \frac{n_2}{2} \geqslant 0
  \, .
\end{equation}

If $V(\lambda)$ is reducible, there exists at least one nontrivial Verma
submodule $V(\lambda')$ in $V(\lambda)$, where $\lambda' = N' \lambda_1$
with $N' < N$.  Equivalently, there are some constants
$a_{jkl} \in \mathbb{C}$, which are not all zero, satisfying
\[
  \sum_{j \geqslant 0} \sum_{k \geqslant 0} \sum_{l \geqslant 0}
    a_{jkl} \, \phi_\lambda^{(jkl)}
  = (\cosh \chi \cos \zeta \, e^{-i \theta})^{N'}
  \, ,
\]
with $0 \leqslant N' < N$.  By virtue of the expression (\ref{eq:f-phi:jkl}),
we have the first observation that $a_{jkl} = 0$ whenever $k \neq l$.
Now that $a_{jll}$ is abbreviated as $a_{j,l}$, we have the second observation
that $a_{j, l} = 0$ whenever $j \neq N - N' - 2 l$.  In the following
$a_{N - N' - 2 l, l}$ is abbreviated as $a_l$.  Then the above condition turns
out to be
\[
  \sum_{l = 0}^{\lfloor \frac{N - N'}{2} \rfloor}
    a_l \, \phi_\lambda^{(N - N' - 2 l, \, l, \, l)}
  = (\cosh \chi \cos \zeta \, e^{- i \theta})^{N'}
  \, ,
\]
namely,
\begin{alignat*}{1}
  &
  \sum_{l = 0}^{\lfloor \frac{N - N'}{2} \rfloor}
    \sum_{j' = 0}^{\lfloor \frac{N - N'}{2} \rfloor - l} \sum_{k' = 0}^l
    (-1)^{N - N' - l + k'} 2^{N - N' - 2 l - 2 j'} \, a_l
    \nonumber \\ & \qquad
    \frac{N!}{(N' + j' + k')!} \,
    \frac{l!}{(l - k')!} \,
    \frac{l!}{k'! \, (l - k')!} \,
    \frac{(N - N' - 2 l)!}{j'! \, (N - N' - 2 l - 2 j')!}
    \nonumber \\ & \qquad
    (\cosh^{N' + 2 l + 2 j'} \chi)
    (\sinh^{N - N' - 2 l - 2 j'} \chi)
    (\cos^{N' + 2 j' + 2 k'} \zeta)
    (1 - \cos^2 \zeta)^{l - k'}
    \, e^{- i \, N' \, \theta}
  \nonumber \\
  =& (\cosh \chi \cos \zeta \, e^{- i \theta})^{N'}
  \, .
\end{alignat*}
Observation of the exponent of $\sinh \chi$ indicates that  $N - N'$ must be
an even integer.  Set $N - N' = 2n$.  Then the above condition becomes
\[
  \sum_{l = 0}^n \sum_{j' = 0}^{n - l} \sum_{k' = 0}^l
    (-1)^{l + k'} 4^{n - l - j'} \, a_l \, b(N, n, j', k', l) \,
    (\cosh^2 \chi)^{l + j'}
    (\cosh^2 \chi - 1)^{n - l - j'}
    (\cos^2 \zeta)^{j' + k'}
    (1 - \cos^2 \zeta)^{l - k'}
  = 1
  \, ,
\]
or, equivalently,
\begin{equation}
  \sum_{j' = 0}^n \sum_{k' = 0}^{n - j'} \sum_{l = k'}^{n - j'}
    (-1)^{l + k'} 4^{n - l - j'} \, a_l \, b(N, n, j', k', l) \,
    (\cosh^2 \chi)^{l + j'}
    (\cosh^2 \chi - 1)^{n - l - j'}
    (\cos^2 \zeta)^{j' + k'}
    (1 - \cos^2 \zeta)^{l - k'}
  = 1
  \, ,
  \label{cond:reducibility}
\end{equation}
where
\[
  b(N, n, j', k', l)
  = \frac{N!}{(N - 2n + j' + k')!}
    \frac{l!}{(l - k')!}
    \frac{l!}{k'! \, (l - k')!}
    \frac{(2 n - 2 l)!}{j'! \, (2 n - 2 l - 2 j')!}
  \, .
\]
We can see from eq.~(\ref{cond:reducibility}) that, for the existence of $a_l$'s
that are not all zero, there must be $n = 0$, namely, $N' = N$.  However, when
$N' = N$, the Verma submodule $V(\lambda') = V(\lambda)$ is no longer trivial.
This proves the irreducibility of the Verma module $V(\lambda)$.

\section{Proof of Eq.~(\ref{eq:multiplicity})}
\label{sect:proof:multiplicity}

In this appendix we prove the formula (\ref{eq:multiplicity}).

Each root $\alpha$ can be associated with a linear transformation
$\sigma_\alpha$ on $\mathfrak{h}'^*$, called a Weyl reflection, sending
$\mu \in \mathfrak{h}'^*$ to $\sigma_\alpha(\mu) \in \mathfrak{h}'^*$, where
\begin{equation}
  \sigma_\alpha(\mu)
  := \mu - \frac{2 \, (\mu, \alpha)}{(\alpha, \alpha)} \, \alpha
  = \mu - \langle \mu, \alpha \rangle \, \alpha
  \, .
\end{equation}
A weight $\mu = n_1 \, \lambda_1 + n_2 \, \lambda_2$ with $n_1 \geqslant 0$ and
$n_2 \geqslant 0$ is called dominant.  Then any weight $\mu \in \Pi(\lambda)$
can be obtained from a dominant weight via a series of Weyl reflections
\cite{Humphreys}: there exist a dominant weight $\mu'$ and some roots
$\beta_1$, \ldots, $\beta_n$ so that
$\mu = \sigma_{\beta_n} \cdots \sigma_{\beta_1}(\mu')$.
Another important fact \cite{Humphreys} is that, for any
$\mu \in \mathfrak{h}'^*$ and any root $\alpha$,
\begin{equation}
  m_\lambda (\sigma_\alpha(\mu)) = m_\lambda(\mu)
  \, .
\end{equation}
It is easy to verify that eq.~(\ref{eq:multiplicity}) does satisfy the above
condition.
The consequence of the above facts is, in order to prove the formula
(\ref{eq:multiplicity}) for arbitrary weight, it is sufficient to prove it
for each dominant weights.

Note that, when $n_1 \, \lambda_1 + n_2 \, \lambda_2$ is a dominant weight,
eq.~(\ref{eq:multiplicity}) turns out to be
\begin{equation}
  m_\lambda(n_1 \, \lambda + n_2 \, \lambda_2)
  = \Big\lfloor \frac{N - n_1 - n_2}{2} \Big\rfloor + 1
  \, .
  \label{eq:multiplicity:dominant}
\end{equation}
We are going to use the Freudenthal formula \cite{Humphreys}
\begin{equation}
  [(\lambda + \delta, \lambda + \delta) - (\mu + \delta, \mu + \delta)]
    m_\lambda (\mu)
  = 2 \sum_{\alpha \in \Phi^+} \sum_{i = 1}^\infty
    (\mu + i \, \alpha, \alpha) \, m_\lambda(\mu + i \, \alpha)
\end{equation}
to prove eq.~(\ref{eq:multiplicity:dominant}) recursively, where
\begin{equation}
  \delta = \frac{1}{2} \sum_{\beta \in \Phi^+} \beta
  = 2 \, (\alpha_1 + \alpha_2) - \frac{1}{2} \, \alpha_1
  \, .
\end{equation}
Note that a weight $\mu = n_1 \, \lambda_1 + n_2 \, \lambda_2$ can be expressed
as $\mu = \lambda - L \, (\alpha_1 + \alpha_2) - l \, \alpha_1$, where
\[
  L := N - n_1 - n_2
  \, , \qquad
  l := \frac{n_2}{2}
  \, .
\]
Then, the recursion is based on $L$ and $l$.

The necessary and sufficient condition for
$\mu = \lambda - L \, (\alpha_1 + \alpha_2) - l \, \alpha_1$
to be dominant is: $L$ and $l$ are integers satisfying
\[
  0 \leqslant L \leqslant N
  \, , \qquad
  0 \leqslant l \leqslant \Big\lfloor \frac{L}{2} \Big\rfloor
  \, .
\]
Furthermore, by virtue of
\begin{alignat}{1}
  \lambda - \mu &= L \, (\alpha_1 + \alpha_2) + l \, \alpha_1
  \, , \\
  \mu + \delta &= (N - L + 2) \, (\alpha_1 + \alpha_2)
  - \Big( l + \frac{1}{2} \Big) \, \alpha_1
\end{alignat}
and the data of inner product in eq.~(\ref{eq:inner-product}), we have
\begin{alignat}{1}
  (\lambda + \delta, \lambda + \delta)
  - (\mu + \delta, \mu + \delta)
  &= (\lambda - \mu, \lambda - \mu)
  + 2 \,(\lambda - \mu, \mu + \delta)
  \nonumber \\
  &= \frac{L}{6} \, (2 N - L + 3)
  + \frac{l}{3} \, (N - L - l + 1)
  \, .
  \label{eq:diff:inner-prod}
\end{alignat}

The Verma module $V(\lambda)$ is generated out of the highest weight $\lambda$.
Hence $m_\lambda(\lambda) = 1$.
If $N = 0$, the only weight is $\lambda = 0$ itself;
If $N = 1$, the only dominant is $\lambda = \lambda_1$ itself.
In these cases eq.~(\ref{eq:multiplicity:dominant}) is obviously correct.
In the following proof we assume that $N \geqslant 2$.

We first recursively prove that a dominant weight
$\mu = \lambda - l \, \alpha_1$ has $m_\lambda(\mu) = 1$.
In fact, when $l = 0$, one has $\mu = \lambda$, for which $m_\lambda(\mu) = 1$
is obviously correct.
As a recursion assumption, we assume that $l > 0$ and that
$m_\lambda(\lambda - i \, \alpha_1) = 1$ for all $i$ satisfying
$0 \leqslant i \leqslant l - 1$.
The Freudenthal formula for $\mu = \lambda - l \, \alpha_1$, then, turns out
to be
\begin{alignat*}{1}
  \frac{l}{3} \, (N - l + 1) \, m_\lambda(\mu)
  &= 2 \sum_{\alpha \in \Phi^+} \sum_{i = 1}^\infty
    (\mu + i \, \alpha, \alpha) \, m_\lambda(\mu + i \, \alpha)
  \, .
\end{alignat*}
For a positive root $\alpha$ and a positive integer $i$,
$\mu + i \, \alpha = \lambda - l \, \alpha_1 + i \, \alpha$ is a weight
(hence $m_\lambda(\mu + i \, \alpha) \neq 0$)
if and only if $\alpha = \alpha_1$, together with
$1 \leqslant i \leqslant l$.  Thus the above equation results in
\begin{alignat*}{1}
  \frac{l}{3} \, (N - l + 1) \, m_\lambda(\mu)
  &= 2 \sum_{i = 1}^{l}
    (\mu + i \, \alpha_1, \alpha_1) \, m_\lambda(\mu + i \, \alpha_1)
  \nonumber \\
  &= 2 \sum_{i = 1}^{l}
    (\lambda - (l - i) \, \alpha_1, \alpha_1) \,
    m_\lambda(\lambda - (l - i) \, \alpha_1)
  \nonumber \\
  &= 2 \sum_{i = 0}^{l - 1}
    (\lambda - i \, \alpha_1, \alpha_1) \, m_\lambda(\lambda - i \, \alpha_1)
  \nonumber \\
  &= 2 \sum_{i = 0}^{l - 1} (\lambda - i \, \alpha_1, \alpha_1)
  \, .
\end{alignat*}
In the last step we have used the recursion assumption
$m_\lambda(\lambda - i \, \alpha_1) = 1$ for $0 \leqslant i \leqslant l - 1$.
The right hand side of the above is
\begin{alignat*}{1}
  2 \sum_{i = 0}^{l - 1} \Big( \frac{N}{6} - \frac{i}{3} \Big)
  = \frac{l}{3} \, (N - l + 1)
  \, .
\end{alignat*}
Hence we have $m_\lambda(\mu) = m_\lambda(\lambda - l \alpha_1) = 1$.
This recursively proves that eq.~(\ref{eq:multiplicity:dominant}) is correct
when $n_1 + n_2 = N$.

Next, for an arbitrary weight $\mu = n_1 \, \lambda_1 + n_2 \, \lambda_2$,
we can observe eq.~(\ref{eq:weight:orthonormal-basis}) and prove that
$N - |n_1 + \frac{n_2}{2}| - |\frac{n_2}{2}|$ is an invariant under the action
of the Weyl group (which is generated by the Weyl reflections).
The meaning of this statement is,
if $w(\mu) = n'_1 \, \lambda_1 + n'_2 \, \lambda_2$ where $w$
is the composition of some Weyl reflections, there will always be the equality
\begin{equation}
  N - \Big| n'_1 + \frac{n'_2}{2} \Big| - \Big| \frac{n'_2}{2} \Big|
  = N - \Big| n_1 + \frac{n_2}{2} \Big| - \Big| \frac{n_2}{2} \Big|
  \, .
\end{equation}
So far, by virtue of the Weyl reflections, we have proved that
eq.~(\ref{eq:multiplicity}) is valid whenever
$N - |n_1 + \frac{n_2}{2}| - |n_2| = 0$.

For convenience, if $\mu = n_1 \, \lambda_1 + n_2 \, \lambda_2
  = l_1 \, \lambda_1 + l_2 \, \alpha_2$
is a weight, we call the invariant
$N - |n_1 + \frac{n_2}{2}| - |\frac{n_2}{2}| = N - |l_1| - |l_2|$
the level of $\mu$.

At last, we make the recursion assumption.  Let $L$ be an integer satisfying
$0 < L \leqslant N$.  We assume that eq.~(\ref{eq:multiplicity}) is valid
whenever $N - |n_1 + \frac{n_2}{2}| - |\frac{n_2}{2}| < L$.  Then we want to
prove that eq.~(\ref{eq:multiplicity:dominant}) is true for dominant weights
of level $L$.  As a conseqence, eq.~(\ref{eq:multiplicity:dominant}) is valid
for all dominant weights, hence eq.~(\ref{eq:multiplicity}) is valid
for all weights.

Since such a dominant weight could be expressed as
$\lambda - L \, (\alpha_1 + \alpha_2) - l \, \alpha_1$,
this proof is recursive by $l$, made up by three steps.

Step 1.  To prove that eq.~(\ref{eq:multiplicity:dominant}) is valid for
the dominant weight $\mu = \lambda - L \, (\alpha_1 + \alpha_2)$.
For this weight, the Freudenthal formula becomes
\begin{alignat}{1}
  \frac{L}{6} \, (2 N - L + 3) \, m_\lambda(\mu)
  &= 2 \sum_{i = 1}^\infty
    (\mu + i \, \alpha_1, \alpha_1) \, m_\lambda(\mu + i \, \alpha_1)
  + 2 \sum_{i = 1}^\infty
    (\mu + i \, \alpha_2, \alpha_2) \, m_\lambda(\mu + i \, \alpha_2)
  \nonumber \\
  &+ 2 \sum_{i = 1}^\infty
    (\mu + i \, (\alpha_1 + \alpha_2), \alpha_1 + \alpha_2) \,
    m_\lambda(\mu + i \, (\alpha_1 + \alpha_2))
  \nonumber \\
  &+ 2 \sum_{i = 1}^\infty
    (\mu + i \, (\alpha_1 + 2 \alpha_2), \alpha_1 + 2 \alpha_2) \,
    m_\lambda(\mu + i \, (\alpha_1 + 2 \alpha_2))
  \, .
  \label{eq:Freudenthal:tmp-0}
\end{alignat}
The summands involve the following weights:
\begin{alignat*}{2}
  \mu + i \, \alpha_1 &= (N - L + i) \, \lambda_1 - i \, \alpha_2
  \, , \qquad &&
  (1 \leqslant i \leqslant \lfloor L/2 \rfloor)
  \, ; \\
  \mu + i \, \alpha_2 &= (N - L) \, \lambda_1 + i \, \alpha_2
  \, , &&
  (1 \leqslant i \leqslant L)
  \, ; \\
  \mu + i \, (\alpha_1 + \alpha_2) &= (N - L + i) \, \lambda_1
  \, , &&
  (1 \leqslant i \leqslant L)
  \, ; \\
  \mu + i \, (\alpha_1 + 2 \alpha_2)
  &= (N - L + i) \, \lambda_1 + i \, \alpha_2
  \, , &&
  (1 \leqslant i \leqslant \lfloor L/2 \rfloor)
  \, .
\end{alignat*}

In fact, when $L = 1$, there exists no weights like $\mu + i \, \alpha_1$ or
$\mu + i \, (\alpha_1 + 2 \alpha_2)$, with $i$ a positive integer.
So this must be specially treated.  In this case $\mu = (N - 1) \, \lambda_1$,
reducing eq.~(\ref{eq:Freudenthal:tmp-0}) to
\[
  \frac{1}{3} \, (N + 1) \, m_\lambda (\mu)
  = 2 \, ((N - 1) \, \lambda_1 + \alpha_2, \alpha_2) \,
    m_\lambda((N - 1) \, \lambda_1 + \alpha_2)
  + 2 \, (N \, \lambda_1, \lambda_1) \, m_\lambda(N \, \lambda_1)
  = \frac{N + 1}{3}
  \, .
\]
Therefore, for $\mu = (N - 1) \, \lambda_1$, we have
$m_\lambda((N - 1) \, \lambda_1) = 1$,
satisfying eq.~(\ref{eq:multiplicity:dominant}).

In the generic case, $1 < L \leqslant N$.
On account of the recursion assumption, eq.~(\ref{eq:Freudenthal:tmp-0})
is reduced to
\begin{alignat}{1}
  \frac{L}{6} \, (2 N - L + 3) \, m_\lambda(\mu)
  &= 2 \sum_{i = 1}^{\lfloor L/2\rfloor} \frac{N - L + 2 i}{3} \, \Big(
      \Big\lfloor \frac{L}{2} \Big\rfloor - i + 1
    \Big)
  + \sum_{i = 1}^L \frac{N - L + 2 i}{3} \,
    \Big( \Big\lfloor \frac{L - i}{2} \Big\rfloor + 1 \Big)
  \nonumber \\
  &= 2 \sum_{i = 1}^{\lfloor L/2 \rfloor} \frac{N - L + 2 i}{3} \, \Big(
      \Big\lfloor \frac{L}{2} \Big\rfloor - i + 1
    \Big)
  + \sum_{i = 0}^{L - 1} \frac{N + L - 2 i}{3} \,
    \Big( \Big\lfloor \frac{i}{2} \Big\rfloor + 1 \Big)
  \, .
  \label{eq:multiplicity:level-L:0}
\end{alignat}
If $L$ is an even number, the above identity turns out to be
\begin{alignat*}{1}
  \frac{L}{6} \, (2 N - L + 3) \, m_\lambda(\mu)
  &= 2 \sum_{i = 1}^{L/2} \frac{N - L + 2 i}{3} \,
    \Big( \frac{L}{2} - i + 1 \Big)
  + \sum_{i = 0}^{L - 1} \frac{N + L - 2 i}{3} \,
    \Big( \Big\lfloor \frac{i}{2} \Big\rfloor + 1 \Big)
  \, .
\end{alignat*}
The second term on the right hand side is
\begin{alignat*}{1}
  & \sum_{i = 0}^{L/2 - 1} \frac{N + L - 4 i}{3} \,
    \Big( \Big\lfloor \frac{2 i}{2} \Big\rfloor + 1 \Big)
  + \sum_{i = 0}^{L/2 - 1} \frac{N + L - 2 \, (2 i + 1)}{3} \,
    \Big( \Big\lfloor \frac{2 i + 1}{2} \Big\rfloor + 1 \Big)
  \nonumber \\
  = & \sum_{i = 0}^{L/2 - 1} \frac{N + L - 4 i}{3} \, ( i + 1 )
  + \sum_{i = 0}^{L/2 - 1} \frac{N + L - 2 \, (2 i + 1)}{3} \, ( i + 1)
  \nonumber \\
  = & \sum_{i = 0}^{L/2 - 1} \frac{2N + 2L - 8 i - 2}{3} \, ( i + 1 )
  = 2 \sum_{i = 1}^{L/2} \frac{N + L - 4 i + 3}{3} \, i
  \, .
\end{alignat*}
Thus, for a dominant weight $\mu = (N - L) \, \lambda_1$ with $L$ an even
integer satisfying $0 < L \leqslant N$,
\begin{alignat*}{1}
  \frac{L}{6} \, (2 N - L + 3) \, m_\lambda(\mu)
  &= 2 \sum_{i = 1}^{L/2} \frac{N - L + 2 i}{3} \,
    \Big( \frac{L}{2} - i + 1 \Big)
  + 2 \sum_{i = 1}^{L/2} \frac{N + L - 4 i + 3}{3} \, i
  \nonumber \\
  &= \frac{L}{6} \, (2 N - L + 3) \, \Big(\frac{L}{2} + 1\Big)
  \, .
\end{alignat*}
It follows that, when $L$ is an even integer satisfying $0 < L \leqslant N$,
\[
  m_\lambda(\mu) = m_\lambda((N - L) \, \lambda_1)
  = \frac{L}{2} + 1
  = \Big\lfloor \frac{L}{2} \Big\rfloor + 1
  = \Big\lfloor \frac{N - |N - L|}{2} \Big\rfloor + 1
\]
satisfies eq.~(\ref{eq:multiplicity}) and eq.~(\ref{eq:multiplicity:dominant}).
If $L$ is an odd number (hence $L \geqslant 3$), in a similar way, we can reduce
eq.~(\ref{eq:multiplicity:level-L:0}) to
\begin{alignat*}{1}
  \frac{L}{6} \, (2 N - L + 3) \, m_\lambda (\mu)
  = \frac{L \, (L + 1)}{12} \, (2 N - L + 3)
  \, .
\end{alignat*}
Therefore, for a dominant weight $\mu = (N - L) \, \lambda_1$
with $3 \leqslant L \leqslant N$ an odd integer,
\[
  m_\lambda (\mu)
  = \frac{L + 1}{2}
  = \Big\lfloor \frac{L}{2} \Big\rfloor + 1
  = \Big\lfloor \frac{N - |N - L|}{2} \Big\rfloor + 1
  \, ,
\]
satisfying eq.~(\ref{eq:multiplicity:dominant}) and
eq.~(\ref{eq:multiplicity}).

So far, we have proved that, the dominant weight
$\mu = \lambda - L \, (\alpha_1 + \alpha_2) = (N - L) \, \lambda_1$,
with $L$ an integer satisfying $0 < L \leqslant N$,
satisfies eq.~(\ref{eq:multiplicity:dominant}) and eq.~(\ref{eq:multiplicity}).
If $L \geqslant N - 1$, there will be no other dominant weight
$n_1 \, \lambda_1 + n_2 \, \lambda_2$ satisfying $N - n_1 - n_2 = L$,
so ends the proof.  In the following, we assume that $0 < L < N - 1$,
provided $N \geqslant 3$.

Step 2.  Let $l$ be an integer satisfying
$0 < l \leqslant \lfloor (N - L)/2 \rfloor$.
We assume that eq.~(\ref{eq:multiplicity:dominant}) is also valid for
dominant weights $\lambda - L \, \lambda_1 - i \, \alpha_1$
with $0 \leqslant i < l$.
We want to prove that eq.~(\ref{eq:multiplicity:dominant}) is also valid for
$\mu = \lambda - L \, \lambda_1 - l \, \alpha_1$.

In fact, the Freudenthal formula for $\mu$ reads
\begin{alignat*}{1}
  & \Big( \frac{L}{6} \, (2 N - L + 3) + \frac{l}{3} \, (N - L - l + 1) \Big)
    \, m_\lambda (\mu)
  \nonumber \\
  = {} & 2 \sum_{i = 1}^{l + \lfloor L/2 \rfloor}
    (\mu + i \, \alpha_1, \alpha_1) \, m_\lambda (\mu + i \, \alpha_1)
  + 2 \sum_{i = 1}^L
    (\mu + i \, \alpha_2, \alpha_2) \, m_\lambda (\mu + i \, \alpha_2)
  \nonumber \\ &
  + 2 \sum_{i = 1}^L
    (\mu + i \, (\alpha_1 + \alpha_2), \alpha_1 + \alpha_2) \,
    m_\lambda (\mu + i \, (\alpha_1 + \alpha_2))
  \nonumber \\ &
  + 2 \sum_{i = 1}^{\lfloor L/2 \rfloor}
    (\mu + i \, (\alpha_1 + 2 \alpha_2), \alpha_1 + 2 \alpha_2) \,
    m_\lambda (\mu + i \, (\alpha_1 + 2 \alpha_2))
  \, .
\end{alignat*}
Using the recursion assumption, we can obtain
\begin{alignat}{1}
  & \Big( \frac{L}{6} \, (2 N - L + 3) + \frac{l}{3} \, (N - L - l + 1) \Big)
    \, m_\lambda (\mu)
  \nonumber \\
  = {} &
  \frac{l}{3} \Big( \Big\lfloor \frac{L}{2} \Big\rfloor + 1 \Big)
    (N - L - l + 1)
  + \Big\lfloor \frac{L}{2} \Big\rfloor
    \Big( \Big\lfloor \frac{L}{2} \Big\rfloor + 1 \Big)
    \Big(
      \frac{N - L}{3}
      + \frac{2}{9} \Big\lfloor \frac{L}{2} \Big\rfloor
      + \frac{4}{9}
    \Big)
  + \frac{(N + 1) \, L}{3}
  \nonumber \\ &
  + \sum_{i = 0}^{L - 1}
    \frac{N + L - 2 i}{3} \, \Big\lfloor \frac{i}{2} \Big\rfloor
  \, .
  \label{eq:Freudental:L-l}
\end{alignat}
When $L$ is an even integer,
\begin{alignat*}{1}
  \sum_{i = 0}^{L - 1}
    \frac{N + L - 2 i}{3} \, \Big\lfloor \frac{i}{2} \Big\rfloor
  &= \sum_{j = 0}^{\frac{L - 2}{2}}
    \frac{N + L - 4 j}{3} \, \Big\lfloor \frac{2 j}{2} \Big\rfloor
  + \sum_{j = 0}^{\frac{L - 2}{2}}
    \frac{N + L - 2 (2j + 1)}{3} \, \Big\lfloor \frac{2 j + 1}{2} \Big\rfloor
  \nonumber \\
  &= \frac{L \, (L - 2)}{4} \, \frac{3 N - L + 1}{9}
  \, .
\end{alignat*}
In this case eq.~(\ref{eq:Freudental:L-l}) results in
\begin{alignat*}{1}
  & \Big( \frac{L}{6} \, (2 N - L + 3) + \frac{l}{3} \, (N - L - l + 1) \Big)
    \, m_\lambda (\mu)
  =
  \Big( \frac{L}{2} + 1 \Big)
    \Big[ \frac{L}{6} \, (2 N - L + 3) + \frac{l}{3} \, (N - L - l + 1) \Big]
  \, ,
\end{alignat*}
namely,
\[
  m_\lambda (\lambda - L \, \lambda_1 - l \, \alpha_1)
  = m_\lambda ((N - L - l) \, \lambda_1 + l \, \alpha_2)
  = \frac{L}{2} + 1
  = \Big\lfloor \frac{L}{2} \Big\rfloor + 1
  = \Big\lfloor \frac{N - |N - L - l| - |l|}{2} \Big\rfloor
  + 1
  \, .
\]
Therefore, when $L$ is even, eq.~(\ref{eq:multiplicity}) is satisfied.
When $L$ is odd,
\begin{alignat*}{1}
  \sum_{i = 0}^{L - 1}
    \frac{N + L - 2 i}{3} \, \Big\lfloor \frac{i}{2} \Big\rfloor
  &= \frac{L - 1}{2} \, \frac{L + 1}{2} \, \frac{3 N - L - 3}{9}
  - \frac{L - 1}{2} \, \frac{N - L}{3}
  \, .
\end{alignat*}
In this case eq.~(\ref{eq:Freudental:L-l}) results in
\begin{alignat*}{1}
  \Big( \frac{L}{6} \, (2 N - L + 3) + \frac{l}{3} \, (N - L - l + 1) \Big)
  \, m_\lambda (\mu)
  = \Big(
      \frac{L}{6} \, (2 N - L + 3) 
      + \frac{l}{3} \, (N - L - l + 1)
    \Big)
    \, \frac{L + 1}{2}
  \, ,
\end{alignat*}
namely,
\[
  m_\lambda (\lambda - L \, \lambda_1 + l \, \alpha_1)
  = m_\lambda ((N - L - l) \, \lambda_1 + l \, \alpha_2)
  = \frac{L + 1}{2}
  = \Big\lfloor \frac{L}{2} \Big\rfloor + 1
  = \Big\lfloor \frac{N - |N - L - l| - l}{2} \Big\rfloor
  + 1
  \, .
\]
Therefore, when $L$ is odd, eq.~(\ref{eq:multiplicity}) is also satisfied.

Step 3. As a summary, we have proved recursively that
eq.~(\ref{eq:multiplicity}) is satisfied for weights of level zero.
Under the recursion assumption that 
weights of level less than $L$ satisfy eq.~(\ref{eq:multiplicity}),
we proved that the weight $\lambda - L \, \lambda_1$ also satisfies
eq.~(\ref{eq:multiplicity}).
Then, under the further recursion assumption that
weights like $\lambda - L \, \lambda_1 - i \, \alpha_1$ with all $i < l$
satisfy eq.~(\ref{eq:multiplicity}), we have proved that dominant
$\lambda - L \, \lambda_1 - l \, \alpha_1$ also satisfy
eq.~(\ref{eq:multiplicity}).
Then, it follows that all weights of level $L$ satisfy
eq.~(\ref{eq:multiplicity}), followed by the final conclusion that all weights
satisfy eq.~(\ref{eq:multiplicity}).



\begin{thebibliography}{99}
  \bibitem{Kay} B.~S.~Kay, ``Quantum field theory in curved spacetime'',
  in \textit{Encyclopedia of Mathematical Physics}, ed. by J.-P.~Fran\c{c}oise,
  G.~Naber and T.~S.~Tsou, Academic (Elsevier) Amsterdam,
  New York and London 2006, Vol.~4 , pp.~202--214;
  arXiv:~\texttt{gr-qc/0601008}.

  \bibitem{Wald06} R.~M.~Wald, ``The History and Present Status of
  Quantum Field Theory in Curved Spacetime'', arXiv:~\texttt{gr-qc/0608018}.

  \bibitem{Hollands-Wald-0803} S.~Hollands and R.~M.~Wald,
  ``Axiomatic quantum field theory in curved spacetime'',
  \textit{Cummun. Math. Phys.} \textbf{293} (2010) 85; arXiv:~0803.2003.

  \bibitem{Hollands-Wald-0805} S.~Hollands and R.~M.~Wald,
  ``Quantum field theory in curved spacetime, the operator product expansion,
  and the dark energy'', \textit{Gen. Rel. Grav.} \textbf{40} (2008) 2051;
  arXiv:~0805.3419.

  \bibitem{Ind-Lib} J.~Indur\'ain and S.~Liberati,
  ``The Theory of a Quantum Noncanonical Field in Curved Spacetimes'',
  \textit{Phys. Rev.} \textbf{D80} (2009) 045008; arXiv:~0905.4568.

  \bibitem{Wald2009} R.~M.~Wald, ``The Forumlation of Quantum Field Theory
  in Curved Spacetime'', arXiv:~0907.0416.



  \bibitem{Wald} See, for example, R.~M.~Wald, \textit{General Relativity},
  the Chicago University Press, 1984.

  \bibitem{Humphreys} J.~E.~Humphreys,
  \textit{Introduction to Lie Algebras and Representation Theory},
  Springer-Verlag, New York 1972.

  \bibitem{Warner} F.~W.~Warner,
  \textit{Foundations of Differentiable Manifolds and Lie Groups},
  Springer-Verlag, New York 1983.

  \bibitem{ChangGuo} Z.~Chang and H.~Y.~Guo, ``Symmetry realization,
  Poisson kernel and the AdS/CFT correspondence'',
  \textit{Mod. Phys. Lett.} \textbf{A15} (2000), 407 (arXiv: hep-th/9910136).

  \bibitem{LHThomas} L.~H.~Thomas,
  ``On Unitary Representations of the Group of de~Sitter Space'',
  \textit{Ann. of Math.} \textbf{42} (1941) 113--126.

  \bibitem{TDNewton} T.~D.~Newton,
  ``A Note on the Representations of the de~Sitter Group'',
  \textit{Ann. of Math.} \textbf{51} (1950) 730--733.

  \bibitem{Wigner} E.~P.~Wigner,
  ``On Unitary Representations of the Inhomogeneous Lorentz Group'',
  \textit{Ann. of Math.} \textbf{40} (1939) 149--204.

  \bibitem{YG} K.~Yagdjian and A.~Galstian, ``Fundamental Solutions for
  the Klein-Gordon Equation in de~Sitter Spacetime'',
  \textit{Comm. Math. Phys.} \textbf{285} (2009), 293--344
  (arXiv: 0803.3074 [math.AP]).

  \bibitem{KV} V.~V.~Kozlov and I.~V.~Volovich, ``Finite Action Klein-Gordon
  Solutions on Lorentzian Manifolds'', \textit{Int. J. Geom. Meth. Mod. Phys.}
  \textbf{3} (2006) 1349--1358 (arXiv: gr-qc/0603111).

  \bibitem{BdS1} H.-Y.~Guo, C.-G.~Huang, Z.~Xu and B.~Zhou,
  ``On Beltrami Model of de~Sitter Spacetime'',
  \textit{Mod. Phys. Lett.} \textbf{A 19} (2004) 1701--1709.

  \bibitem{BdS2} H.-Y.~Guo, C.-G.~Huang, Z.~Xu and B.~Zhou,
  ``On special relativity with cosmological constant'',
  \textit{Phys. Lett.} \textbf{A 311} (2004) 1--7.

  \bibitem{BdS3} H.-Y.~Guo, C.-G.~Huang, Y.~Tian, H.-T.~Wu, Z.~Xu and B.~Zhou,
  ``Snyder's model---de~Sitter special relativity duality and
    de~Sitter gravity'',
  \textit{Class. Quantum Grav.} \textbf{24} (2007) 4009--4035.

  \bibitem{BdS4}
  H.-Y.~Guo, C.-G.~Huang, Y.~Tian, Z.~Xu and B.~Zhou,
  ``Snyder's quantized space-time and de~Sitter special relativity'',
  \textit{Front. Phys. China} \textbf{2} (2007) 358--363.
\end{thebibliography}
\end{document}